\documentclass[prd,aps,superscriptaddress,twocolumn,nofootinbib,showpacs,preprintnumbers,floatfix]{revtex4}

\pdfoutput=1

\usepackage{graphicx}
\usepackage{dcolumn}
\usepackage{amsfonts,amsmath,amssymb,amsthm}
\usepackage{euscript,bbm}
\usepackage{ifthen}
\usepackage{psfrag}

\usepackage{subdepth}

\usepackage{comment}

\usepackage{bm}


\usepackage{slashed}

\declareslashed{}{/}{0}{-0.05}{E}
\declareslashed{}{/}{0}{-0.05}{P}
\declareslashed{}{/}{0}{-0.10}{\Widehat{P}}

\usepackage{float}
\usepackage{fancybox}
\usepackage{fancyvrb}
\usepackage{ragged2e}
\cornersize*{15pt}
\newfloat{card}{htb}{lop}
\floatname{card}{Card}

\makeatletter
\renewcommand\floatc@plain[2]{\setbox\@tempboxa\hbox{{\footnotesize {\bf {\@fs@cfont #1:}} #2}}%
\ifdim\wd\@tempboxa>\hsize {\begin{minipage}[t]{0.9\linewidth}\justifying \small {\raggedleft {\@fs@cfont #1:} #2}\end{minipage}}\par
\else\hbox to\hsize{\hfil\box\@tempboxa\hfil}\fi}
\@namedef{thecard}{\Alph{card}}
\makeatother

\def\met{{\slashed{E}} {}_{\rm T}}

\def\fsu5{\mbox{$\cal{F}$-$SU(5)$}}
\def\bfsu5{$\boldsymbol{\mathcal{F}}$-$\boldsymbol{SU(5)}$}

\newcommand{\be}{\begin{equation}}
\newcommand{\ee}{\end{equation}}
\newcommand{\bea}{\begin{eqnarray}}
\newcommand{\eea}{\end{eqnarray}}

\newcommand{\gsim}{\lower.7ex\hbox{$\;\stackrel{\textstyle>}{\sim}\;$}}
\newcommand{\lsim}{\lower.7ex\hbox{$\;\stackrel{\textstyle<}{\sim}\;$}}

\newcommand{\ttbar}{t \bar{t}}

\newcommand{\pT} {{P_{\rm T}}}

\newcommand{\PlotPairWide}[4]{
\begin{figure*}[htp]
\centering
\hspace{-5pt}
\includegraphics[width=3.4in]{#1}
\hspace{+10pt}
\includegraphics[width=3.4in]{#2}
\begin{minipage}[l]{0.8\linewidth}\vspace{-10pt} \justifying \small {\raggedleft \caption{#4} \label{#3}}\end{minipage}\vspace{0pt}
\end{figure*}}

\newcommand{\PlotSingle}[4]{
\begin{figure}[htp]
\centering
\includegraphics[width=#2]{#1}
\begin{minipage}[l]{0.8\linewidth}\vspace{-0pt} \justifying \small {\raggedleft \caption{#4} \label{#3}}\end{minipage}\vspace{0pt}
\end{figure}}

\begin{document}

\title{Third Generation in Cascade Decays}

\author{Bhaskar Dutta}
\affiliation{George P. and Cynthia W. Mitchell Institute for Fundamental Physics and Astronomy, Texas A$\&$M University, College Station, TX 77843, USA}
\author{Tianjun Li}
\affiliation{State Key Laboratory of Theoretical Physics and Kavli Institute for Theoretical Physics China (KITPC),
Institute of Theoretical Physics, Chinese Academy of Sciences, Beijing 100190, P. R. China}
\affiliation{School of Physical Electronics, University of Electronic Science and Technology of China,
Chengdu 610054, P. R. China }
\author{James A. Maxin}
\affiliation{Department of Physics and Engineering Physics, The University of Tulsa, Tulsa, OK 74104 USA}
\author{Dimitri V. Nanopoulos}
\affiliation{George P. and Cynthia W. Mitchell Institute for Fundamental Physics and Astronomy, Texas A$\&$M University, College Station, TX 77843, USA}
\affiliation{Astroparticle Physics Group, Houston Advanced Research Center (HARC), Mitchell Campus, Woodlands, TX 77381, USA}
\affiliation{Academy of Athens, Division of Natural Sciences, 28 Panepistimiou Avenue, Athens 10679, Greece}
\author{Kuver Sinha}
\affiliation{Department of Physics, Syracuse University, Syracuse, NY 13244, USA}
\author{Joel W. Walker}
\affiliation{Department of Physics, Sam Houston State University, Huntsville, TX 77341, USA}

\begin{abstract}
In supersymmetric models with gluinos $m({\tilde g}) \sim 1000-2000$ GeV, new physics searches based on cascade decay products
of the gluino are viable at the next run of the LHC.
We investigate a scenario where the light stop is lighter than the gluino and both are lighter than all other squarks,
and show that its signal can be
established using multi $b$-jet, multi $W$ and/or multi lepton final state topologies. We then utilize both boosted
and conventional jet topologies in the final state in conjunction with di-tau production as a
probe of the stau-neutralino co-annihilation region responsible for the model's dark matter content.
This study is performed in the specific context of one such phenomenologically viable model named No-Scale \fsu5. 
\end{abstract}
MIFPA-14-36;
ACT-12-14
\maketitle


\section{Introduction\label{sct:intro}}


Weak-scale supersymmetry (SUSY) provides a leading candidate for physics beyond the Standard Model (SM), as it addresses the hierarchy problem,
gives gauge coupling unification, and (in $R$-parity conserving models) provides a robust dark matter (DM) candidate. 
The search for colored superpartners at the Large Hadron Collider (LHC) has so far yielded null results. The exclusion
limits on squark ($\tilde{q}$) and gluino ($\tilde{g}$) masses, when they are comparable, are approximately $1.5$ TeV at $95\%$ CL with
$20$ fB$^{-1}$ of integrated luminosity \cite{Aad:2012fqa, Aad:2012hm, Chatrchyan:2012lia, TheATLAScollaboration:2013fha}.
If the squarks are heavy, the gluino mass is constrained to be heavier than approximately 1.2 TeV at $95\%$ CL with
$20$ fB$^{-1}$ of integrated luminosity \cite{Aad:2012fqa, Aad:2012hm, Chatrchyan:2012lia, TheATLAScollaboration:2013fha, CMS:2014dpa}.

There are many interesting scenarios where the squarks are heavy but the gluino mass is about a TeV or so. In this paper we probe
one such scenario, i.e., No-Scale \fsu5~\cite{Li:2013naa,Maxin:2011hy}.  In this model, the Higgs mass receives supplementary
contributions from additional TeV scale vector-like fields~\cite{Li:2011ab}, which may make it easier to obtain a Higgs mass
near 126 GeV without tuning in the stop sector.
The light stop $\tilde{t}_1$ is lighter than the gluino in this model,
and by an amount that allows for on-shell decays with unity branching ratio for most of the viable parameter space.
Similar types of mass spectra may also occur in large volume scenario models~\cite{Aparicio:2014wxa}.
Once gluinos are produced, they may cascade via the stop into the (Wino-like) second lightest neutralino or light chargino, which then decay into the lightest neutralino via stau,
e.g. $\tilde\chi^0_2\rightarrow\tilde\tau_1\tau\rightarrow\tau\tau\tilde\chi^0_1$. The potentially sizeable mass gap between the first and
second neutralino will tend to impart a sizeable kinematic boost
to the lead tau production, but the narrow conduit $\Delta M \equiv M_{{\tilde{\tau}}_{1}} - M_{{\tilde{\chi}}^0_1}$
linking the stau and LSP may render the secondary tau rather soft, and hence more difficult to detect,
barring inheritance of a substantive upstream boost.  The dark matter content may be obtained thermally
by stau and (Bino-like) light neutralino coannihilation in the case of a small $\Delta M \simeq 6$~GeV.

The final state of such a cascade contains $b$-quarks, $\tau$ leptons, missing transverse energy $\met$, and also
additional light flavor leptons.  The scenario can be established by the presence of multiple $b$-quarks and leptons in the final state.
Establishing the coannihilation element would require $b$'s and $\tau$'s in the final state, where $t\bar t$ is the dominant
background for such a signal, also containing $b$'s, $\tau$'s and missing energy.
In this paper we first try to establish the scenario from the multi-$b$, leptons and $W$'s, so that the existence of a third
generation can be surmised.  Similar final state event topologies have been studied recently at the LHC~\cite{CMS:2014dpa}.
We then try to establish the signal in the coannihilation region by considering two analysis routes, each of which requires
at least two $\tau$'s with $\pT > (40,20)$ GeV in $|\eta|<2.5$ and large missing energy in the final state,
but with distinct handling of the jets. In the scenario that we will refer to as the ``Boosted Event Topology'',
two tagging jets $j_{1,2}$ with $\pT > (75,50)$~GeV in $|\eta| \leq 5.0$ are required, with 
an opposite hemisphere orientation $(\eta_1\times\eta_2 < 0)$, and an absolute separation in pseudo-rapidity of $|\eta_2-\eta_1|>3.5$.
In the scenario that we will refer to as the ``Conventional Event Topology'', the two leading jets $j_{1,2}$ in  $|\eta| < 2.5$
carry very large transverse momenta $\pT > (400,200)$~GeV.  No vetoes on heavy flavor jets or light lepton flavors are
enforced.  We will demonstrate the cut flow optimization for all the selections, and calculate the signal to background
ratios S/B.  A methodology using the boosted event topology has been used in the context of stop searches recently~\cite{Dutta:2013gga} and
is found to be very effective for background reduction.  A variation of conventional topology  has been prescribed in~\cite{Arnowitt:2008bz} with the
requirement that the jets have to be non-$b$ jet to reduce the background.  In our case study, however, the signal contains a large
population of $b$-jets arising from decays of the on-shell stop, which makes controlling the $t\bar{t}$
background for establishing the coannihilation region more challenging.

The structure of the paper is as follows. In Section \ref{sct:model}, the model context is described for this work, followed by event generation and selection in Section
\ref{sct:generation}.  The primary results and analysis are presented in Sections \ref{sct:signal} (establishment of the signal),
\ref{sct:ditau} (di-tau event topologies), and \ref{sct:mtt} (resolving the neutralino-stau system).
Conclusions are given in Section \ref{sct:conclusions}.


\section{No-Scale \bfsu5 Model Context\label{sct:model}}

In order to provide a specific context for long-chain
cascades of the described type arising from the gluino decay when the squarks are heavy, this study will reference a class of well-defined models named No-Scale \fsu5
(see Ref.~\cite{Li:2013naa} and references therein), which combine $(i)$ field content of the
Flipped $SU(5)$ grand unified theory (GUT), with $(ii)$ a pair of hypothetical TeV-scale vector-like supermultiplets (``flippons'')
of mass $M_V$ derivable within local F-Theory model building, and $(iii)$ the boundary conditions of No-Scale Supergravity (SUGRA).

The signature collider characteristic of this model family is a light
stop $t_1$, lighter than the gluino $\tilde{g}$, which is in turn lighter than all other squarks $\tilde{q}$.
Lightness of the gluino is attributable to flippon-induced modifications to evolution of the renormalization group equations (RGEs),
specifically a nullification at first-loop of the color-associated beta-function coefficient $(b_3=0)$; these modifications
simultaneously elevate the scale of secondary Flipped $SU(5)\times U(1)_{\rm X}$ gauge unification into adjacency with $M_{\rm Planck}$,
allowing for phenomenologically consistent implementation of the notoriously delicate No-Scale framework~\cite{Ellis:2001kg,Ellis:2010jb}.
The gluino to stop plus top decay $\tilde{g} \rightarrow \tilde{t}\overline{t}$ is on-shell,
with a 100\% branching ratio in the viable parameter space, and is associated with an extreme prevalence of hadronic jets~\cite{Maxin:2011hy}.

The key phenomenological characteristic of this model is
proportional rescaling at leading order of the full model spectrum with respect to only a single high scale dimensionful
parameter, the unified gaugino mass $M_{1/2}$.  The model is only logarithmically sensitive to the specific threshold scale $M_V$ at
which the vector-like multiplets actively circulate in loops. Lower order facility is available to exchange small variations of $\tan \beta$, the
ratio of up- and down-like Higgs vacuum expectation values, within a narrow range around $20-25$ for a corresponding
fluctuation in the stau-LSP mass gap $\Delta M$ between about 5 and 25 GeV.  The neutralino LSP dark matter candidate is always dominantly Bino.

This model thereby provides a tightly constrained, yet realistic, testbed for a study of the described type, which
may moreover exemplify a formally smooth transition between thermal and non-thermal dark matter scenarios.
The \fsu5 SUSY spectrum for $M_{1/2} \sim 1$~TeV and $\Delta M \sim 6$~GeV is provided in Table~\ref{bench1}.
This example is in the stau-neutralino coannihilation region, with thermal Bino dark matter providing the observed relic density.
The selected mass range is in the vicinity of the exclusion boundary established data from the 7 and 8~TeV Large Hadron Collider (LHC) runs;
commencement of collisions near the 14~TeV design energy will actively probe the \fsu5 construction at scales above $M_{1/2} = 1$~TeV. 
The technology to be described is generically applicable for all UV models
that feature light gluinos, for example those based alternatively on the $SO(10)$ GUT, and can be employed broadly in the next run of LHC.

\begin{table}[ht]
  \small
    \caption{Spectrum (in GeV) for $M_{1/2} = 990$~GeV, $M_{V} = 8044$~GeV, $m_{t} = 174.4$~GeV, and $\tan \beta$ = 23.3.
	Here, $\Omega_{\rm CDM} h^2$ = 0.1197, the stau-LSP mass gap is $\Delta M = 6.4$~GeV, and the lightest neutralino is greater than 99\% Bino.
        For other values of $M_{1/2}$, revisions to the complete SUSY spectrum may be very well approximated by a simple proportional rescaling.  $\Delta M$
        may be increased by slightly lowering $\tan \beta$, with minimal additional effect on the spectrum overall.}
\hspace{-5pt}\begin{minipage}{\linewidth}
\begin{tabular}{|c|c||c|c||c|c||c|c||c|c||c|c|} \hline
     $\widetilde{\chi}_{1}^{0}$ & $213$ & $\widetilde{\chi}_{1}^{\pm}$ & $449$ & $\widetilde{e}_{R}$ &    $366$ & $\widetilde{t}_{1}$ & $1104$ & $\widetilde{u}_{R}$ & $1824$ & $m_{h}$ &         $125.1$\\ \hline
     $\widetilde{\chi}_{2}^{0}$ & $449$ & $\widetilde{\chi}_{2}^{\pm}$ & $1463$ & $\widetilde{e}_{L}$ &   $989$ & $\widetilde{t}_{2}$ & $1672$ & $\widetilde{u}_{L}$ & $1985$ & $m_{A,H}$ &       $1590$\\ \hline
     $\widetilde{\chi}_{3}^{0}$ & $1461$ & $\widetilde{\nu}_{e/\mu}$ &   $986$ & $\widetilde{\tau}_{1}$ & $220$ & $\widetilde{b}_{1}$ & $1650$ & $\widetilde{d}_{R}$ & $1887$ & $m_{H^{\pm}}$ &   $1592$\\ \hline
     $\widetilde{\chi}_{4}^{0}$ & $1463$ & $\widetilde{\nu}_{\tau}$ &    $958$ & $\widetilde{\tau}_{2}$ & $964$ & $\widetilde{b}_{2}$ & $1789$ & $\widetilde{d}_{L}$ & $1986$ & $\widetilde{g}$ & $1328$\\ \hline
	\end{tabular}
\end{minipage}
\label{bench1}
\end{table}

\section{Event Generation and Selection\label{sct:generation}}

Numerical analysis of the parameter interdependencies in No-Scale \fsu5 is conducted with {\tt SuSpect~2.34}~\cite{Djouadi:2002ze},
utilizing a proprietary codebase modification that incorporates the flippon-enhanced RGEs.
Signal and standard model (SM) background Monte Carlo event samples,
including parton showering and fast detector simulation, are generated via the standard
{\sc MadGraph5}/{\sc MadEvent}~\cite{Alwall:2011uj}, {\sc Pythia}~\cite{Sjostrand:2006za}, {\sc PGS4}~\cite{PGS4} chain.
Subclassifications of SUSY two-body production channels, such as $\tilde{g}\tilde{g}$, $\tilde{g}\tilde{q}$, $\tilde{q}\tilde{q}^{*}$, and etc.,
are individually simulated for a variety of \fsu5 benchmark configurations with 0, 1, or 2 hard jets (generated inclusively) at the matrix element level.
There is some approximation here (at a level consistent with other technological and systematic limitations) associated with (for example) the
reembedding of certain diagrams for squark production with 0 jets into diagrams for gluino pair production with 2 jets.
{\sc MadEvent} is configured, in conjunction with {\sc Pythia}, to use MLM matching with kt jets in the ``Shower kT'' scheme, setting
({{\tt ickkw}} $\Rightarrow$ {{\tt 1}}),
({{\tt xqcut}} $\Rightarrow$ {{\tt 200}}),
({{\tt drjj}} $\Rightarrow$ {{\tt 0}}),
({{\tt auto\_ptj\_mjj}} $\Rightarrow$ {{\tt T}}),
({{\tt QCUT}} $\Rightarrow$ {{\tt 200}}), and
({{\tt SHOWERKT}} $\Rightarrow$ {{\tt T}}).
The {\sc PGS4} detector simulation employs a standard LHC-appropriate parameter card, with jet clustering performed
using the anti-kt algorithm.  Given the vital importance of hadronic tau reconstruction to this study, the somewhat
dated ``shrinking cone'' algorithm native to PGS4 has been replaced by a customized treatment that applies a flat
60\% detection efficiency (with a 1\% fake rate) to tau objects itemized at the {\sc Pythia} level; the invisible
tau neutrino 4-momentum is appropriately subtracted.  The algorithms for tagging of heavy flavor jets have also
been customized, with curves selected for an efficiency around 70\%, and the acceptance region in pseudo-rapidity
for generic jet candidates has been extended to $|\eta| < 5.0$.
Selection cuts are implemented on the detector-level event simulation within {\sc AEACuS~3.6}~\cite{Walker:2012vf,aeacus}.
Initial event selections for the various targeted final states are
made very conservatively, to allow for subsequent optimization.

The strongest signal of new physics in a $m(\tilde{q}) > m(\tilde{g}) > m({\tilde{t}}_1)$ type model
is expected in association with extremely long cascade decay chains, featuring a strong
four $W$ plus four $b$ heavy flavor jet component~\cite{CMS:2014dpa}.
Since the $W$ may decay leptonically ($1/3$ for three light generations) or
hadronically ($2/3$ for two light generations times 3 colors), the final state will also be profused
with leptons and multi-jets.  In order to establish the signal, we
therefore require at least two $b$-jets in all cases,
while recording the net count of jets, leptons, di-leptons, and missing transverse energy $\met$,
expecting $(i)$ that events with fewer leptons should have more jets, and $(ii)$ that
the dominant $\ttbar+ {\rm Jets}$ background may likewise have large jet counts, but should not
generally feature very large $\met$ values. 
For $t\overline{t} + {\rm Jets}$, charge conservation further implies that any dilepton production
must be anti-correlated in sign, whereas the independent leptonic decay
events are uncorrelated in flavor. The SUSY four $W+b$ signal may readily
produce tri-leptons (category III), which are inaccessible, outside of fakes, to
$t\overline{t} + {\rm Jets}$; this category, which necessarily 
includes also a like-sign dilepton, should be intrinsically low background.
Likewise, the orthogonal categorization of precisely two leptons (category II)
with like sign should intrinsically suppress $t\overline{t} + {\rm Jets}$,
with residual fakes, sign-flips, etc., reduced by a requirement on
missing transverse energy $\met$.  The remaining event subdivisions (category I),
i.e. those with $0,1,~{\rm or}~2$ leptons, but no like-sign dilepton,
will rely heavily on the missing energy cut for background reduction,
but may also feature a much stronger net signal count.  Opposite sign
di-tau production, which is subsequently studied in detail, is a
subset of this very broad event category (though the final jet requirements
will ultimately be quite different). 

In the $2j + 2\tau + \met$ final state, which is vital to probing the question of
dark matter coannihilation, we preliminarily select on 
$(i)$ two or more isolated taus with $\pT > (40,20)$ GeV in $|\eta| < 2.5$;
$(ii)$ two jets $j_{1,2}$ with $\pT > 20$~GeV in $|\eta| < 2.5$ for the conventional scenario, {\it OR}
two jets $j_{1,2}$ with $\pT > 20$~GeV, $(\eta_1 \times \eta_2 < 0)$, and $|\eta_2-\eta_1|>3.5$ in $|\eta| < 5.0$ for the boosted scenario.
More specifically, in the latter scenario, the pair satisfying these criteria that has the largest invariant mass is tagged,
and additional jets interior to this pair are allowed.
The {\sc AEACuS} instructions (with comments) applicable to these selections (and to the computation and storage of other
analysis parameters such as the $\met$, jet kinematics, heavy flavor and light lepton counts, di-tau sign orientation,
transverse momentum components, and invariant mass) are documented in Card~\ref{CARD:GENERAL}.

\begin{card}[htp]
\centering
\ovalbox{%
\begin{minipage}{0.88\linewidth}
\vspace{3pt}
{\scriptsize
\begin{Verbatim}[numbers=left,numbersep=6pt]
******** cut_card.dat v3.6 ***
* DiTau, Jets and MET Searches
*** Object Reconstruction ****
OBJ_JET     = PTM:20, PRM:[0.0,2.5]
  # Jet candidates with P_T > 20 & |ETA| < 2.5
  # OBJ_JET     = PTM:20, PRM:[0.0,5.0]
  # Alternative "VBF" Wide Jets for boosted scenario
OBJ_JET_001 = SRC:+000, CUT:[2,UNDEF,-1]
  # Classifies Leading Jet Pair objects
  # OBJ_JET_001 = SRC:+000, EFF:[-1,3.5,UNDEF,1],
  # CUT:[2,UNDEF,-1] # Alternative "VBF" Topology
  # for boosted scenario with Delta ETA > 3.5
OBJ_JET_002 = SRC:+001, CUT:[1,UNDEF,-1],
  OUT:[PTM_001,MAS_001,ETA_001,PHI_001]
  # Outputs Lead Jet kinematics
OBJ_JET_003 = SRC:[+001,-002],
  OUT:[PTM_002,MAS_002,ETA_002,PHI_002]
  # Outputs Soft Jet kinematics
OBJ_JET_004 = SRC:[+002,+003], HFT:0.5, CUT:0
  # Counts Heavy Flavor Jets (no cut)
OBJ_LEP     = PTM:10, PRM:[0.0,2.5]
  # Soft Leptons restricted in P_T and ETA 
OBJ_LEP_001 = SRC:+000, PTM:20, EMT:+3, SDR:[0.3,UNDEF,1],
  CUT:[2,UNDEF,-1] # Two isolated Taus with P_T > 20
OBJ_LEP_002 = SRC:+001, PTM:40, CUT:[1,UNDEF,-1],
  OUT:PTM_003 # Lead Tau has P_T > 40
OBJ_LEP_003 = SRC:[+001,-002], OUT:PTM_004
  # Object holds the Soft Tau (no cut)
OBJ_LEP_004 = SRC:+000, EMT:-3, CUT:0
  # Counts soft light leptons (no cut)
OBJ_DIL_001 = LEP:001, DLS:+1, CUT:0
  # Counts Like Sign DiTaus (no cut)
OBJ_DIL_002 = LEP:001, DLS:-1, CUT:0
  # Counts Opposite Sign DiTaus (no cut)
****** Event Selection *******
EVT_MET     = CUT:0
  # Outputs Missing Energy (no cut)
EVT_OIM_001 = LEP:001, CUT:0
  # Outputs DiTau Invariant Mass (no cut)
******************************
\end{Verbatim}
} \vspace{-3pt} \end{minipage}}
\caption{{\sc AEACuS} instruction card for di-Tau, jets and missing transverse energy $\met$ searches.}
\label{CARD:GENERAL}
\end{card}

\section{Establishing the \bfsu5 Signal\label{sct:signal}}

\begin{card}[htp]
\centering
\ovalbox{%
\begin{minipage}{0.88\linewidth}
\vspace{3pt}
{\scriptsize
\begin{Verbatim}[numbers=left,numbersep=6pt]
******** cut_card.dat v3.6 ***
* B-Jets, Jets, Leptons, and MET Searches
*** Object Reconstruction ****
OBJ_JET     = PTM:20, PRM:[0.0,2.5]
  # Jet candidates with P_T > 20 & |ETA| < 2.5
OBJ_JET_001 = SRC:+000, PTM:40, CUT:8
  # Force 8 Jets with P_T > 40
OBJ_JET_002 = SRC:+001, PTM:80, CUT:4
  # Force 4 Jets with P_T > 80
OBJ_JET_003 = SRC:+002, PTM:200, CUT:2
  # Force 2 Jets with P_T > 200 
OBJ_JET_004 = SRC:+003, PTM:400, CUT:1
  # Force 1 Jet with P_T > 400
OBJ_JET_005 = SRC:+002, HFT:0.5, CUT:2
  # Force 2 b-Jets with P_T > 80 
OBJ_LEP     = PTM:20, PRM:[0.0,2.5]
  # Soft Leptons restricted in P_T and ETA 
OBJ_LEP_001 = SRC:+000, SDR:[0.3,UNDEF,1], CUT:[0,2]
  # Force 0, 1, or 2 Leptons
OBJ_DIL_001 = LEP:001, DLS:+1, CUT:[0,0]
  # Veto like sign dileptons
****** Event Selection *******
EVT_MET     = CUT:700
 # Cut on Missing Energy below 700 GeV 
******************************
\end{Verbatim}
} \vspace{-3pt} \end{minipage}}
\caption{{\sc AEACuS} instruction card for the Category I event selection with
$b$-jets, jets, leptons, and missing transverse energy $\met$.  The example
may be readily modified for the Category II or III event selections.}
\label{CARD:FINAL_STATE}
\end{card}

\PlotSingle{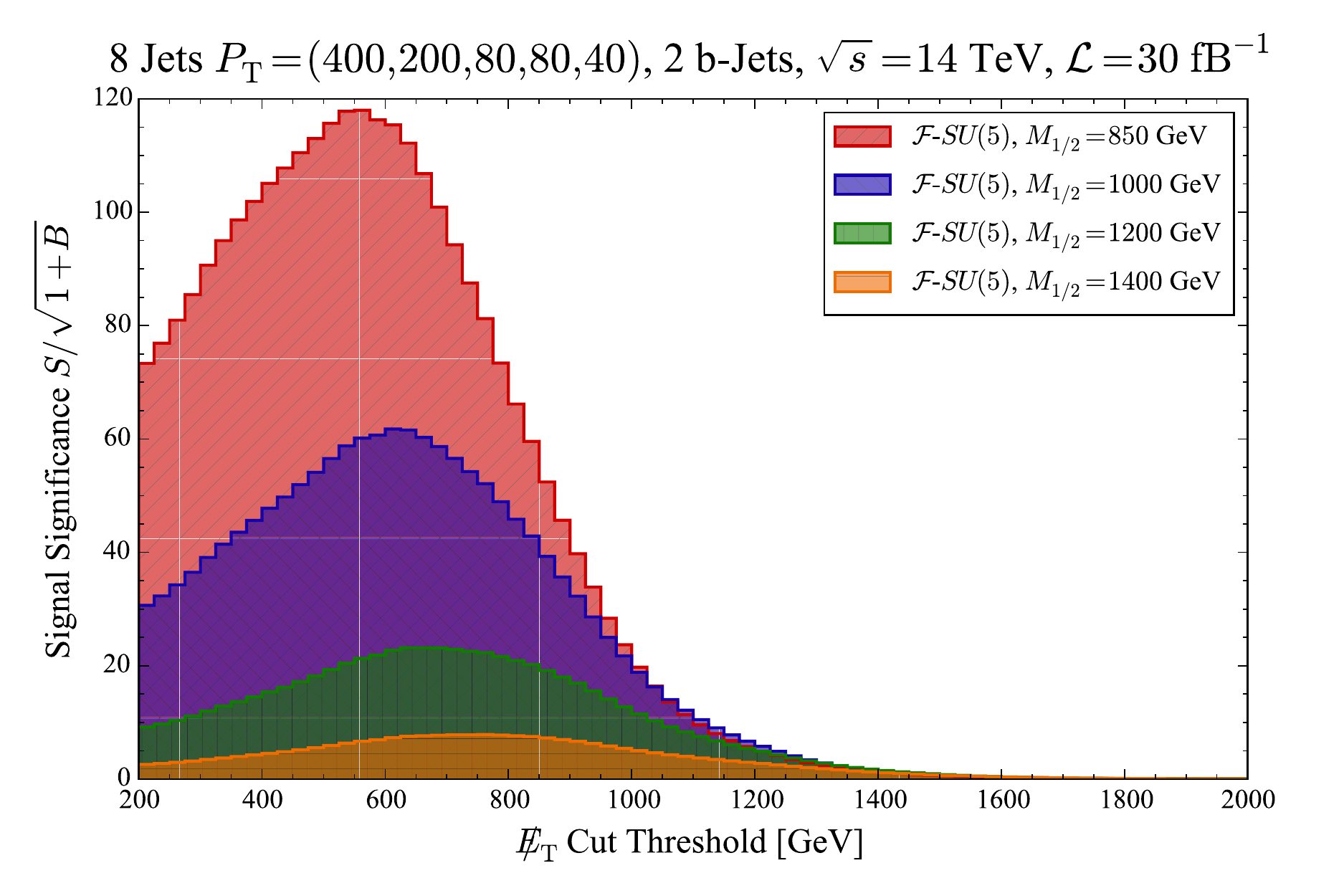}{3.4in}{fig:catI}
{Signal significance is evaluated as a function of the
missing transverse energy $\met$ cut for events with (0,1,2) leptons and no
 like-sign dilepton (category I) at a luminosity of 30 events per femtobarn.
Two heavy-flavor tagged jets with $P_{\rm T} > 80$~GeV are required, and the leading
eight jets (with or without a b-tag) must carry $P_{\rm T} > (400,200,80,80,40)$~GeV.}

\PlotPairWide{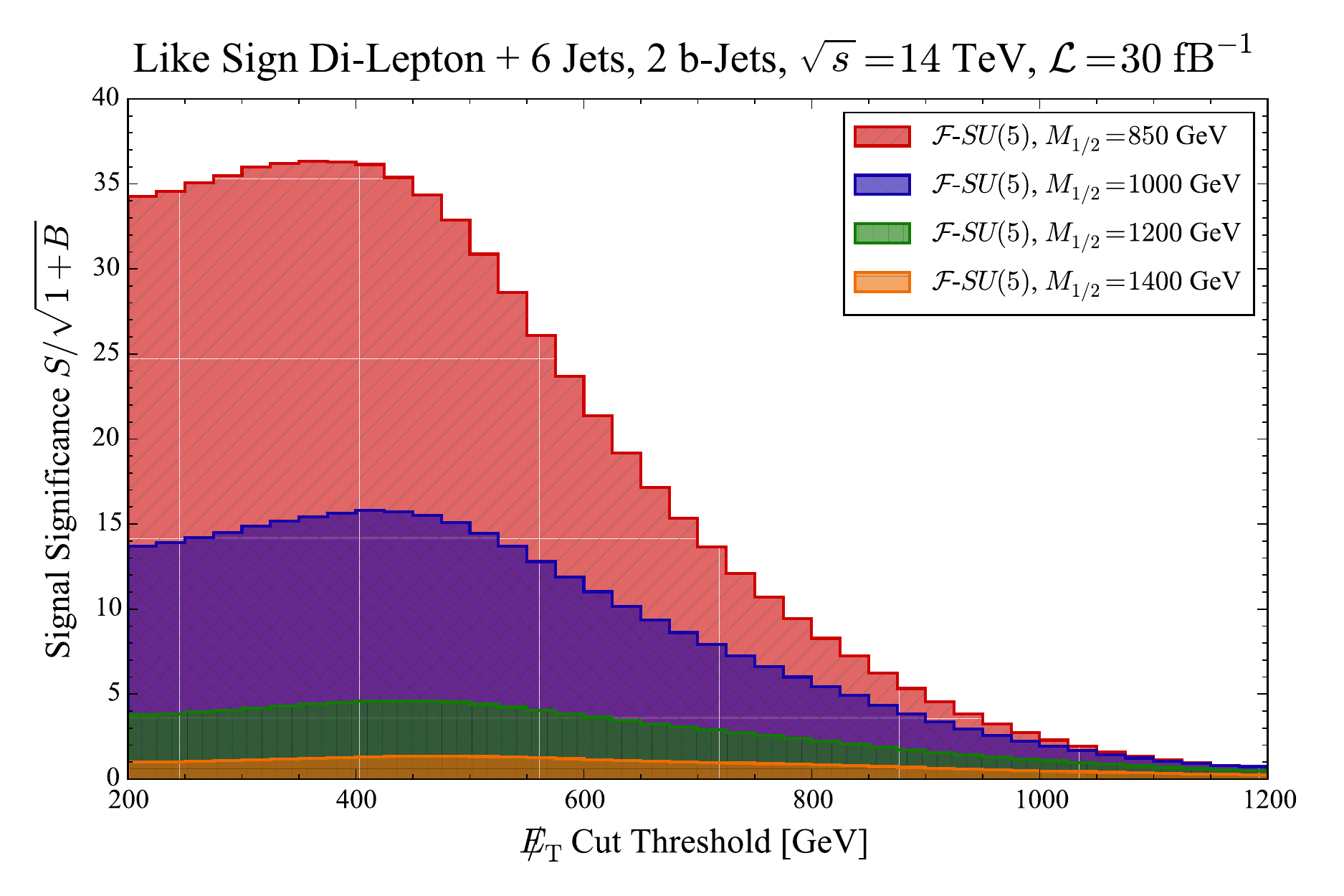}{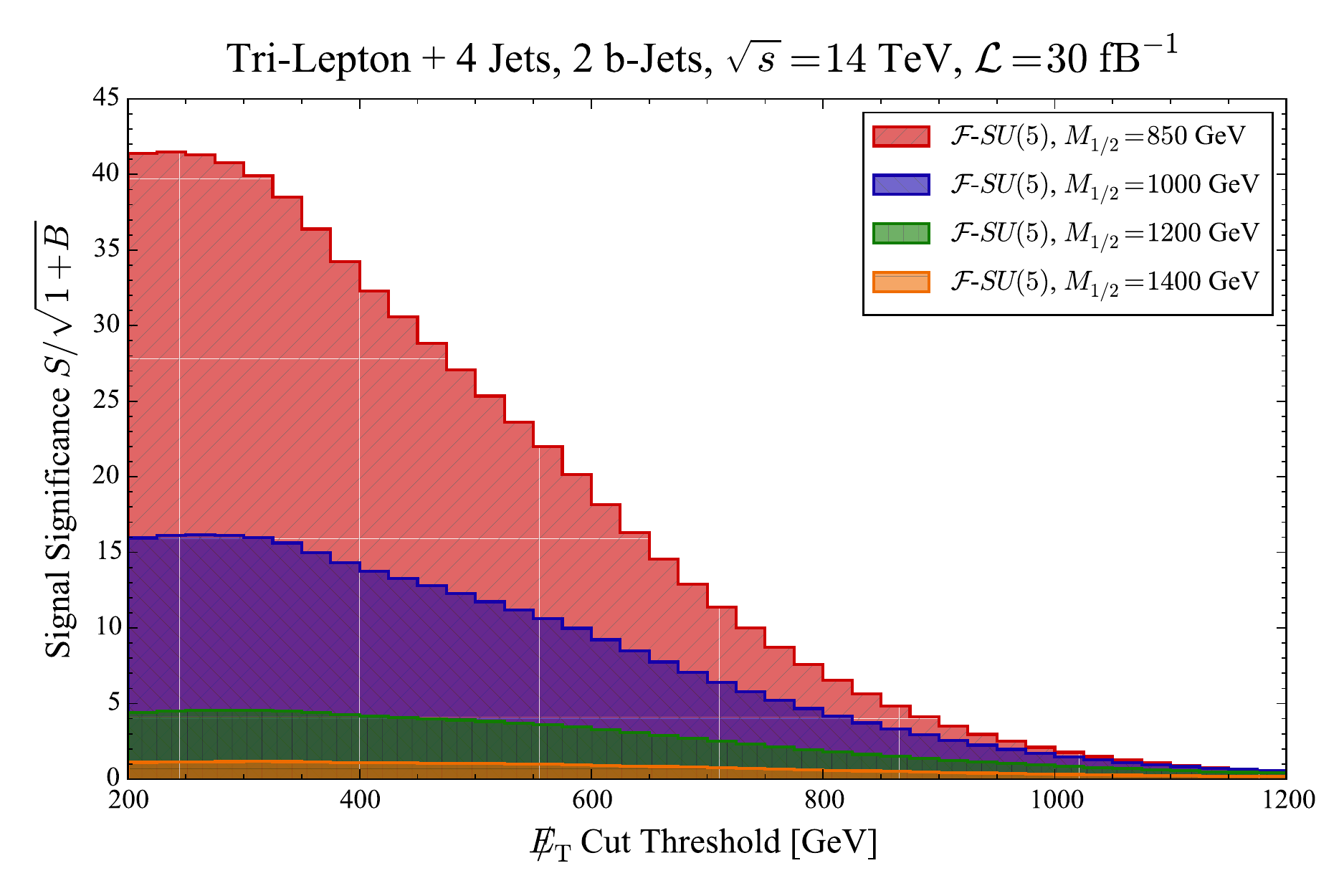}{fig:catII-III}
{Signal significance is evaluated as a function of the
missing transverse energy $\met$ cut for the like-sign dilepton (category II) and tri-lepton
(category III) event topologies at a luminosity of 30 events per femtobarn.
Two heavy-flavor tagged jets with $P_{\rm T} > 80$~GeV are required, and the leading
four jets (with or without a b-tag) must carry $P_{\rm T} > (400,200,80,80)$~GeV.
For the like-sign dilepton categorization, two additional jets (for a total of 6)
must carry $P_{\rm T} > 40$~GeV.}

\PlotPairWide{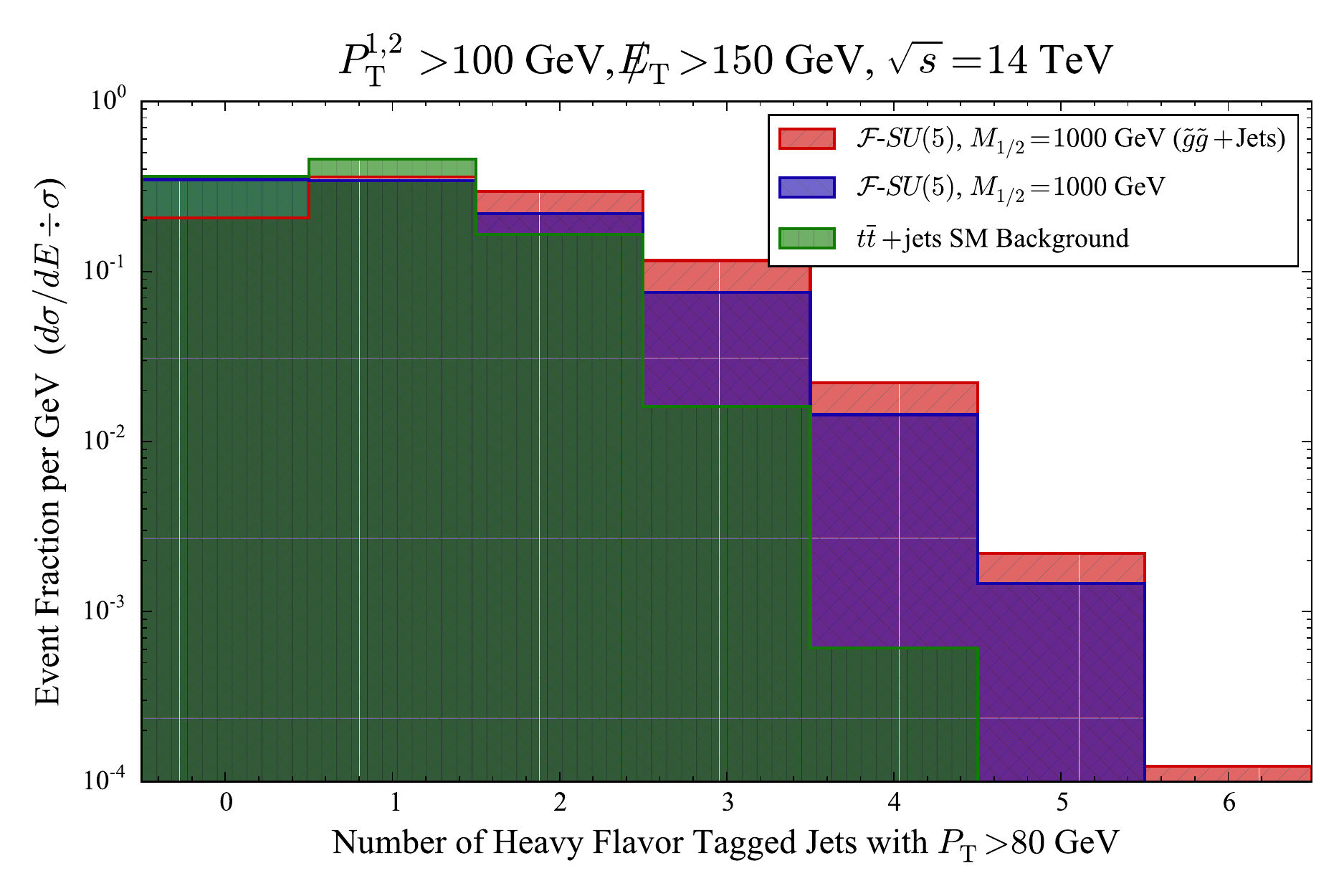}{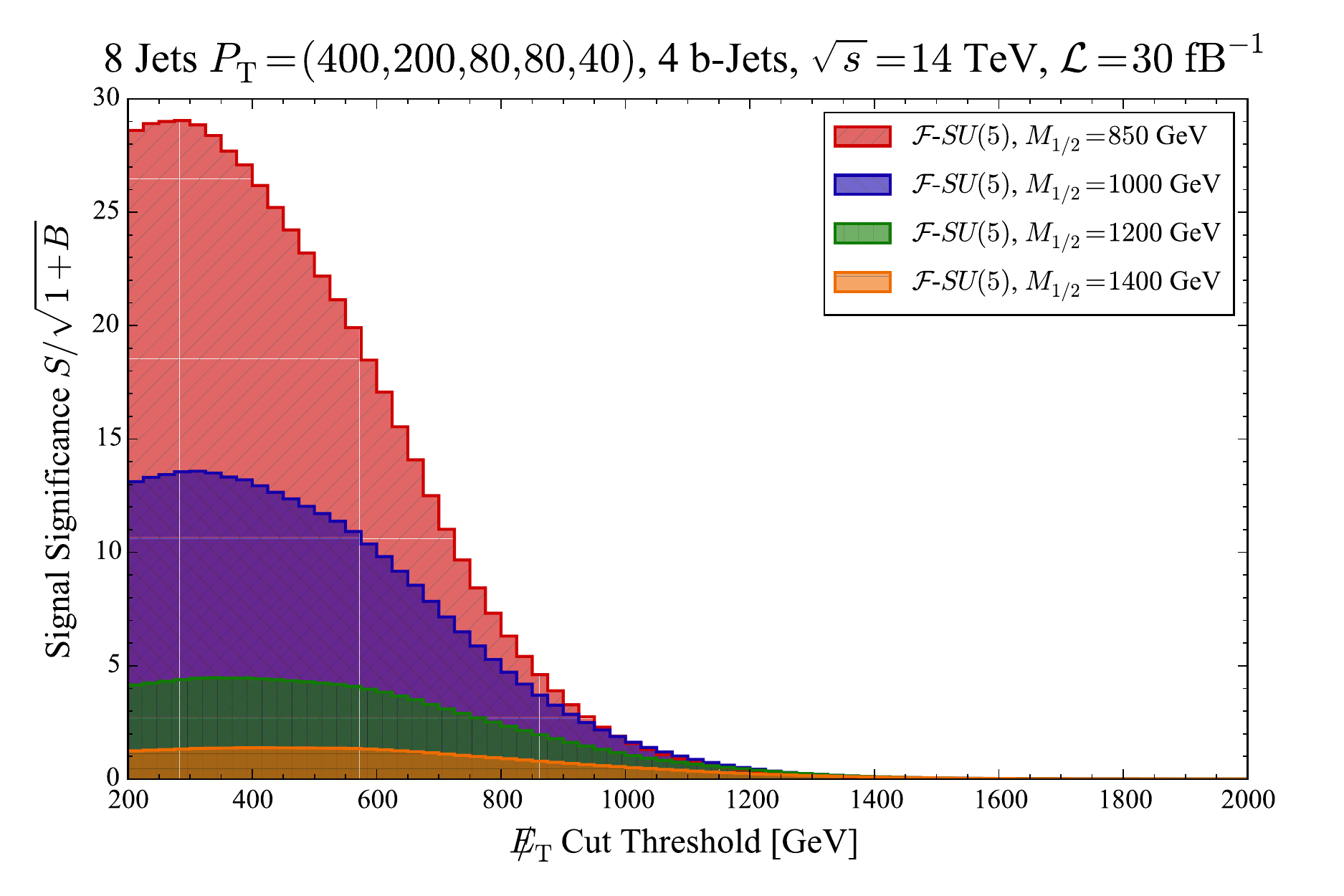}{fig:b}
{Distribution of $b$-jets retained after soft kinematic grooming in the background,
the unified signal, and the isolated gluino pair production channel (left).  The signal
significance is evaluated as a function of the missing transverse energy $\met$ for the
Category I selections, adding a 4 $b$-jet requirement relative to the selection
depicted in Fig.~(\ref{fig:catI}).}

\PlotSingle{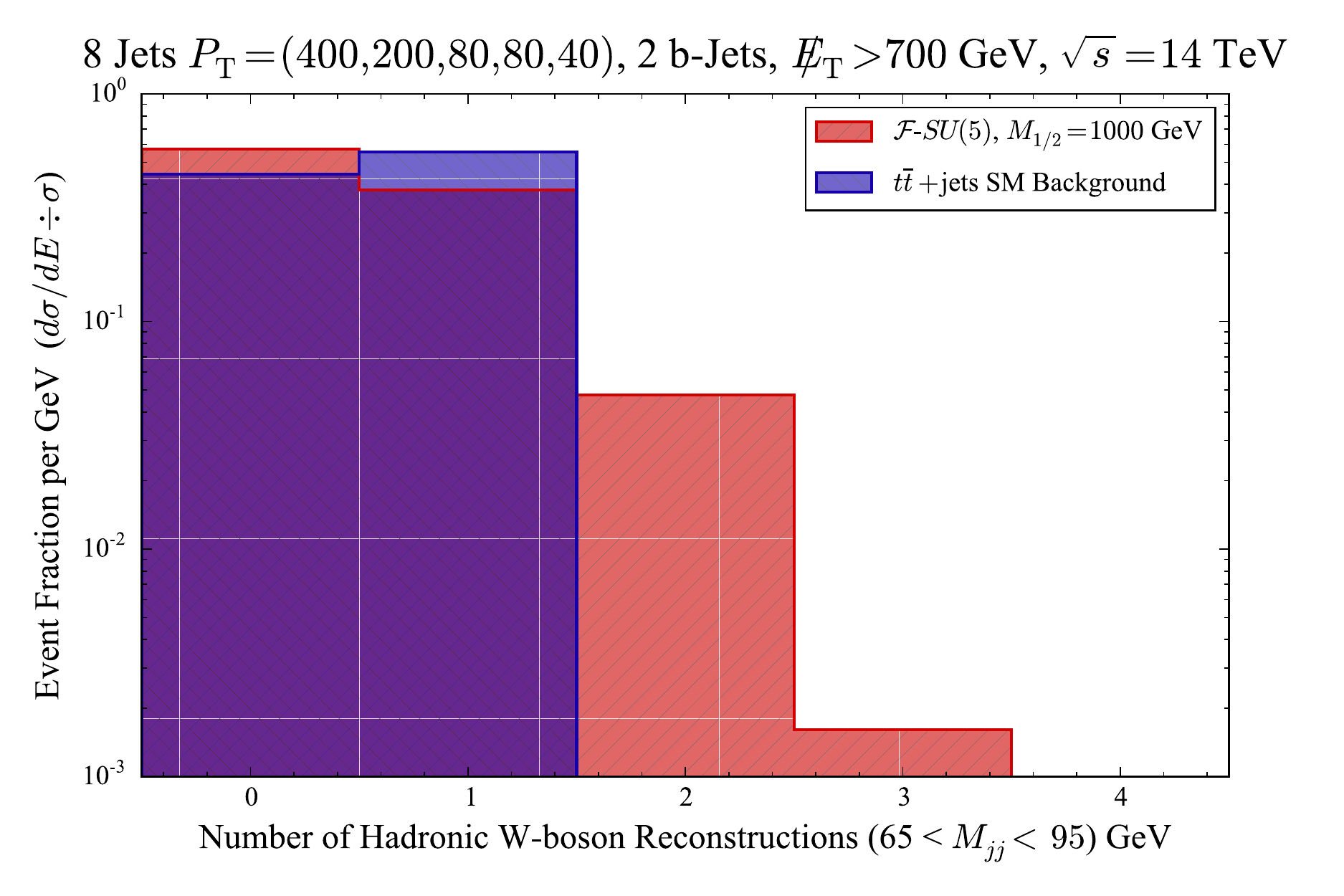}{3.4in}{fig:W}
{Distribution of reconstructed hadronic $W$'s in the Category I $\geq$ 8 jets with $\geq$ 2 $b$-jets
plus large $\met$ final state for both signal and background.}

The described 2 $b$-jet signal categories (I,II,III) have been established in Monte Carlo simulation.
The results are little changed for a $b$-jet transverse momentum threshold of
$40$~GeV versus $80$~GeV, so the more robust value of 80 is retained.  The intuition that something
like $8,6,4$ hard jets may be respectively expected in the signal for each category
is well confirmed, noting that the latter categories exchange
jet pair production for lepton production in the $W$ decay.  Any jets associated
with a squark to gluino transition (typically a 500 GeV to 750 GeV mass gap)
are expected to be quite hard.  Jets downstream from the stop decay also
receive a substantial boost from the mass differential, and all downstream
jets may inherit large kinematic boosts, even in decays with less phase space.
Requiring $P_{\rm T}^{1,2} > (400,200)$~GeV facilitates very robust tagging
on the leading jet pair, while dampening background (allowing a lower $\met$ floor),
and retaining excellent signal statistics.  Jets 3 and 4 are well resolved at
$P_{\rm T}^{3,4} > 80$~GeV, consistent with the $b$-jet threshold, whereas any jets
required beyond the leading four are better captured with softer threshold
around $P_{\rm T}^{5+} > 40$~GeV.  With these cuts in place, the missing transverse
energy $\met$ threshold may be individually optimized for each category,
as demonstrated in Figs.~(\ref{fig:catI},\ref{fig:catII-III}).
We will select $\met > (700,500,300)$~GeV, respectively.
The {\sc AEACuS} instructions (with comments) applicable to the Category I final state selections
are documented in Card~\ref{CARD:FINAL_STATE}, and may be readily adapted for application
to categories II or III.

The background is found in each case to be extraordinarily well controlled,
with excellent signal retention.  Categories III (tri-leptons) and II
(like-sign di-leptons) appear to be observable up to about $M_{1/2} \sim 1200$~GeV,
while the primary category I (all other events) may be probed beyond $M_{1/2} \sim 1400$~GeV,
encompassing the majority, if not totality, of the \fsu5 model space.  The gluino
masses in these cases are on the order of 1600 and 1900 GeV, respectively. As
demonstrated clearly in Fig.~(\ref{fig:catI}), the expected SUSY event yield is
a strongly decaying function of $M_{1/2}$, which may be inverted in order to
establish the global \fsu5 mass scale.  Since the model is dominantly single
parameter, the bulk properties of the spectrum are then fixed, and may be
cross-correlated against alternatively designed event selections for consistency,
such as the di-tau production channel to be elaborated subsequently.

In Fig.~(\ref{fig:b}), we show that the signal is indeed more rich in heavy flavor jets than the
background, even after considering $b$-fake rates.  It should be recalled that these rates also
receive a sizeable contribution from the hadronic decays of $W$ to charm/strange.  We also show
the significance optimization for various $M_{1/2}$ values in $\geq$ 8 jets +$\geq 4$  $b$-jets + $\met$ final states.
It is clear, in comparison to Fig.~(\ref{fig:catI}), that the stronger $b$-jet requirement allows for 
a softer $\met$ cut, although this strategy is not necessarily a more favorable one.

In Fig.~(\ref{fig:W}), we depict the density of hadronic $W$-bosons reconstructed
with an invariant mass inside the 65 to 95~GeV window out of non-$b$ tagged jets,
after application of all the Category I cuts.
In this limit, the signal is showing more $W$ counts, although a trend toward
under representing the expected prevalence of $W$'s has been observed.
In the gluino pair production channel, at $M_{1/2} \sim 1$~TeV, two top/stop pairs are produced,
and the stop will produce a top/neutralino with probability $p \sim 0.86$, or a bottom/chargino with probability
$(1-p) \sim 0.14$, such that the expected density of top quarks is
$2_t\times{(1-p)}^2 + 3_t\times2\,p\,{(1-p)} + 4_t\times{p}^2 \sim 3.7$, each of which
should be associated with a final state $W$-boson.  This has been validated in Monte Carlo at the generator level,
although is apparently difficult to resolve at the detector level using elementary techniques.  Several contributions to the
$W$ shortage are trivial, including exclusion of leptonic branching (the neutrino cannot be deconvolved from the large SUSY
$\met$ content), the likelihood of heavy flavor tagging for decays to charm/strange, limits on kinematic acceptance of jets,
and smearing of the jet energy resolution.  However, difficulty in clearly distinguishing the SUSY hadronic $W$ shape distribution
from that of $\ttbar$ suggests that a more subtle agent is also at work.  Specifically, it would seem to implicate the absorption
of distinct partonic chains into combined fat jets, as is made more likely in a cascade environment with extremely high
jet density, and also when decay products are highly boosted (collimated);  it is likely that a jet substructure
analysis~\cite{Khachatryan:2014gha,Khachatryan:2014hpa} would improve discrimination, but this will not be considered further here.

\PlotSingle{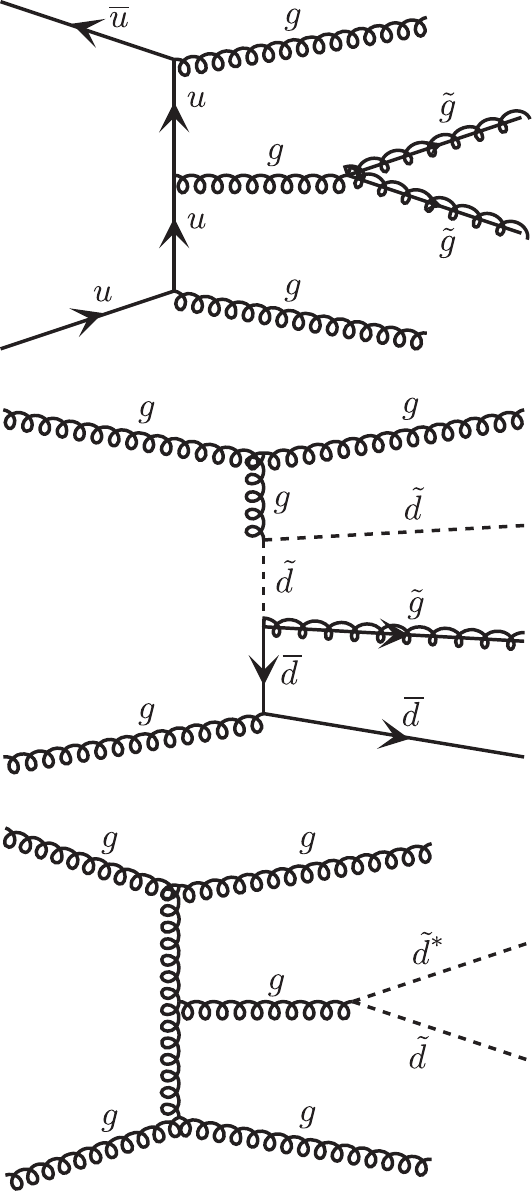}{2.0in}{Fig:STRONG}
{Feynman diagrams (adapted from {\sc MadGraph5}~\cite{Alwall:2011uj} output) contributing to strong interaction analogs of the
electroweak ``Vector Boson Fusion'' boosted event topology with di-jet plus 
$\tilde{g}\tilde{g}$, $\tilde{g}\tilde{q}$, and $\tilde{q}\tilde{q}^{*}$ final states.}

\section{Di-Tau Event Results and Analysis\label{sct:ditau}}

Having established a strong third generation signal component in the prior section, we now
turn attention to the question of stau/LSP coannihilation, via a focus on di-tau event production.
Vector boson fusion (VBF)~\cite{Dutta:2012xe,Delannoy:2013ata,Delannoy:2013dla} is a key electroweak process
capable of highlighting the light neutralino system at the four-vertex order. 
The VBF process is inherently suppressed,
but may be emphasized by imposing an event topology that features highly boosted tagging jets with wide separation in pseudorapidity.
Also, there are a large variety of strongly-interacting diagrams which may be considered VBF analogs that may
likewise give rise to the described event topology, such as those exhibited in Fig.~(\ref{Fig:STRONG}).
Such diagrams cannot directly invoke the neutralino system, but may still probe the electroweak sector downstream, generating similar di-tau production
via long-chain decay cascades, in association with a (potentially substantially) elevated net count of jets.
By strength of the interaction and of numbers, such processes may residually swamp their electroweak analogs.

\PlotPairWide{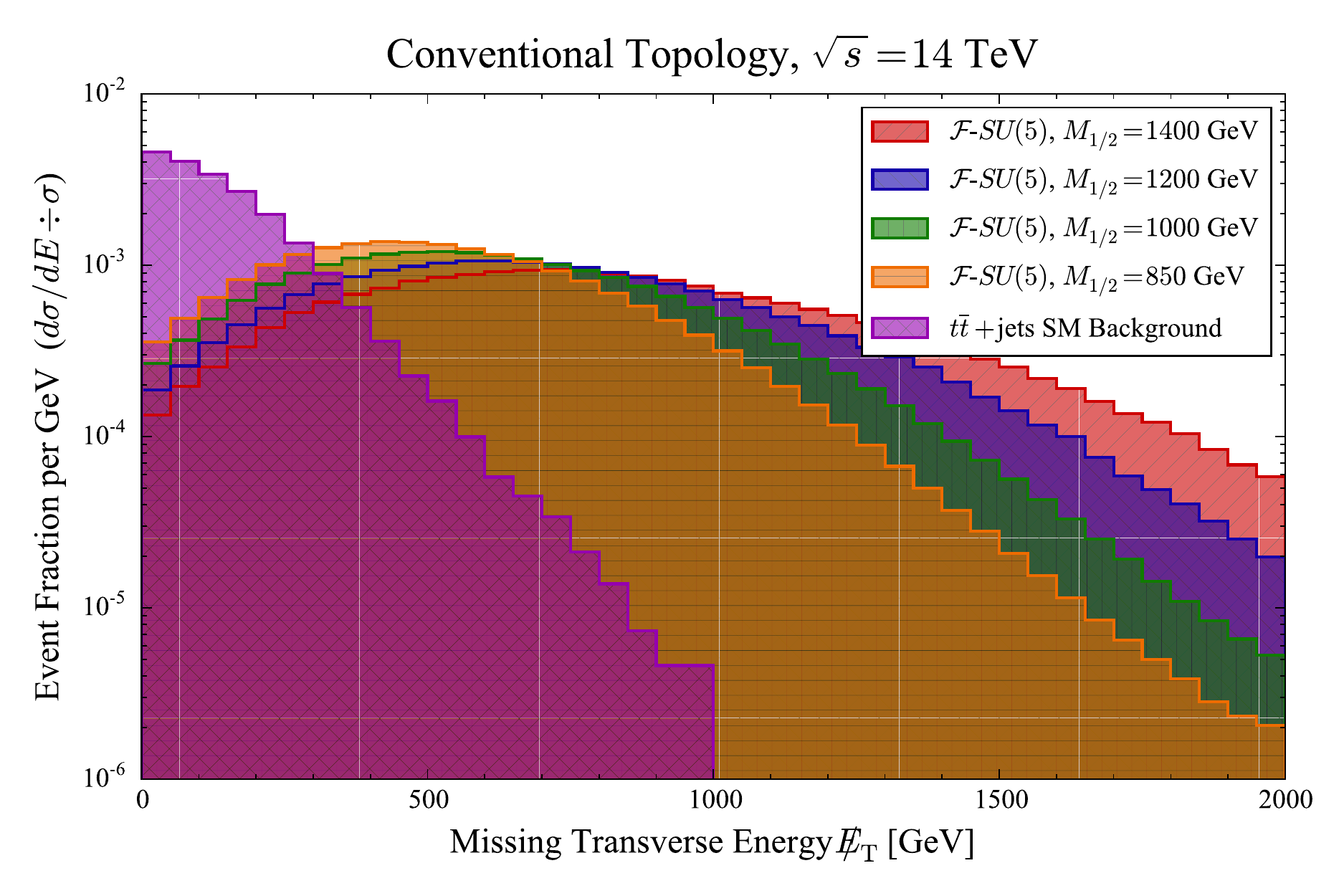}{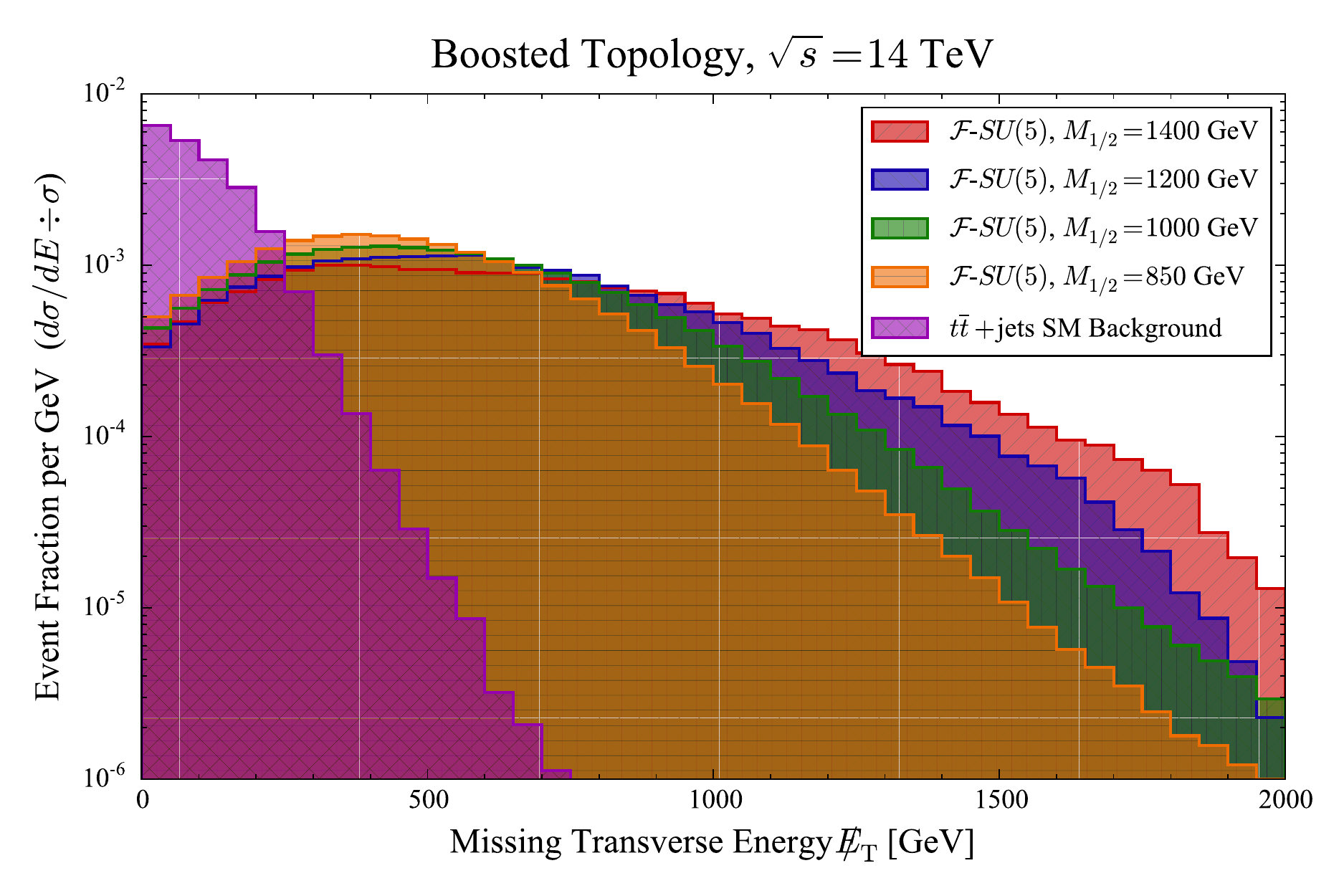}{Fig:met_shape}
{Normalized signal and background distribution shapes are compared as a function of the missing
transverse energy $\met$ for the boosted and conventional event topologies.}

\PlotPairWide{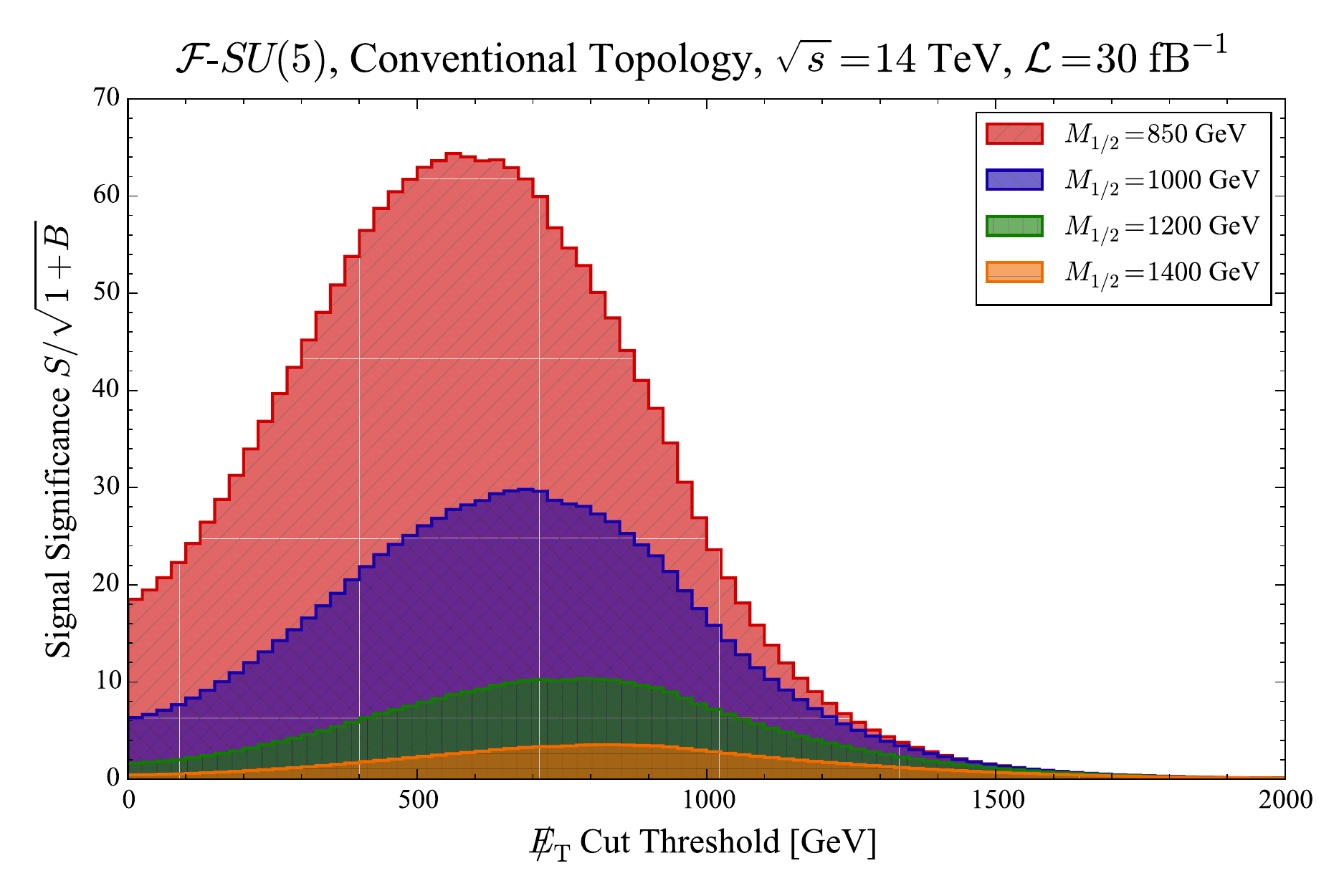}{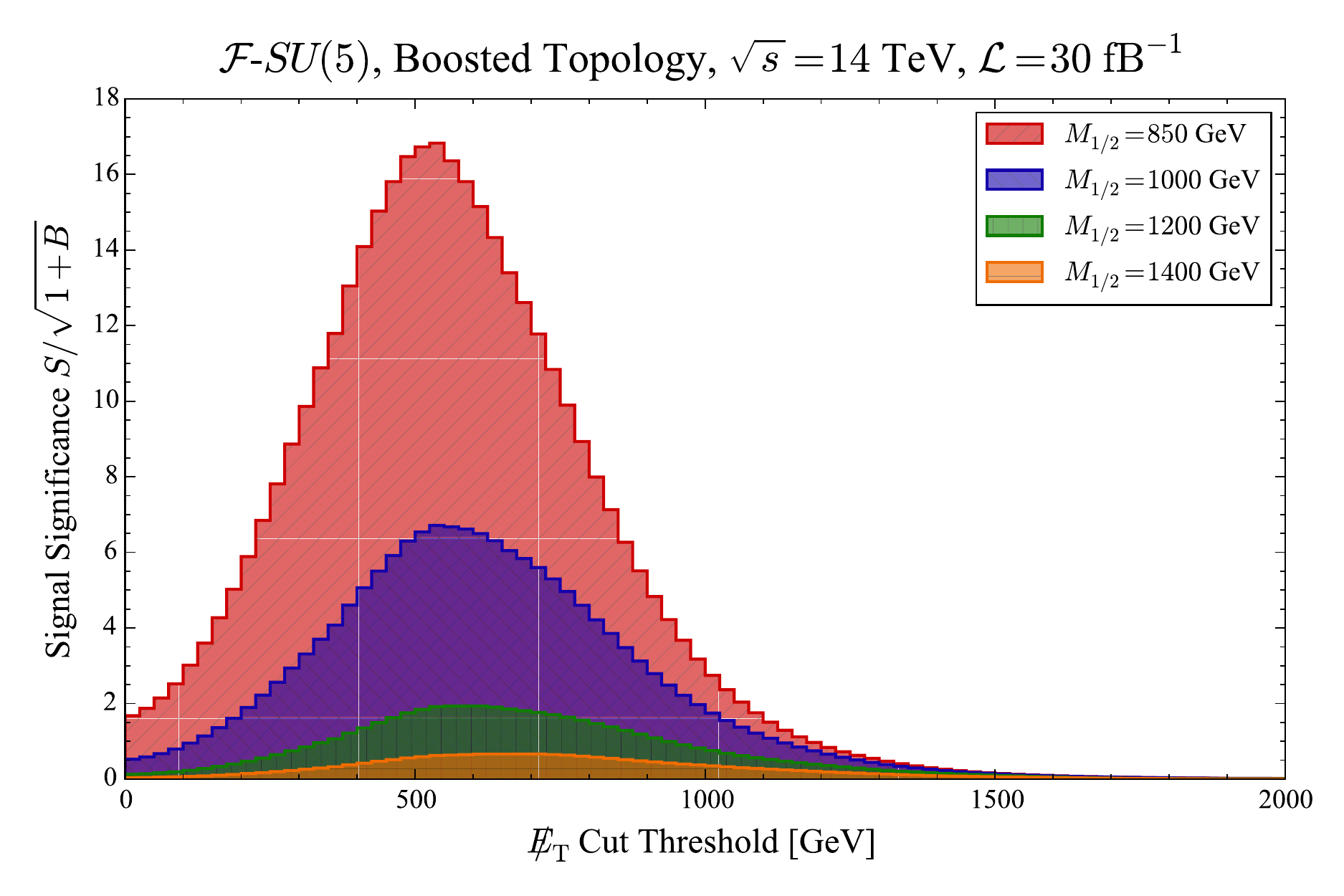}{Fig:met_sig}
{Signal significance is evaluated as a function of the
missing transverse energy $\met$ cut for the boosted and conventional event topologies
at a luminosity of 30 events per femtobarn.  All other di-tau and jet cuts documented in
Section~\ref{sct:intro} are imposed as a prerequisite.}

The signal significance metric $S/\sqrt{1+B}$ has been optimized as a function of the missing energy and
transverse momenta of the two relevant jets for each of the two considered event topologies (boosted/conventional),
with and without the application of light lepton and heavy flavor jet vetoes.  This optimization,
which is mildly luminosity and scale dependent, has been performed at an integrated luminosity of 30 events
per fB, taking an \fsu5 model benchmark with $M_{1/2} = 1000$~GeV and $\Delta M = 25$~GeV.
Including a lepton veto hurts the significance ratio by about 30\% for both event topologies.
Including a $b$-jet veto hurts the boosted case by about 10\%, but the conventional case by 20 or 30\%.
This latter difference may be traced in part to the fact that the boosted topology employs 
jets at large $\eta$ that cannot be tagged for heavy flavor, such that the veto is less harmful.
Both results are consistent with the expectation that long cascades down the squark to gluino
to stop to neutralino cascade may fork a large population of both light leptons and $b$-jets.
The possibility of widening the jet pseudorapidity acceptance to $|\eta| < 5.0$ for the conventional 
event topology has also been examined, and found to be inconsequential with regards both to net signal efficiency
and the optimization structure; this observation may be traced to the fact that the two leading jets
in $P_{\rm T}$ are not likely to exist at large pseudorapidity, as these conditions are anti-correlated.

For both scenarios, the most important
cut is on missing energy, and values around
700~GeV work very well.  In the conventional 
event topology, the signal significance at
$\mathcal{L} = 30/{\mathrm fB}$ is moderately,
though not greatly, enhanced by the addition of cuts
on the leading jet transverse momenta $P_{\rm T}$.
This is because the background would already be well controlled
by the $\met$ cut alone and the "+1" regulator term is playing a deciding
role in the denominator.  Nevertheless signal
is not much harmed by pushing the jet momenta substantially
(a large $\met$ will require large jet momenta)
and background is thereby suppressed in a much more
robust manner.  This is supported by the Table~\ref{tab:flow}
cut flow, presented in terms of the residual fB cross section.
Specifically, the addition of a cut $P_{\rm T} > (400,200)$
reduces the background by about 150 times, but the
signal by only about 2.5 times;  this is prior to
implementation of any $\met$ cut whatsoever.

\bgroup
\def\arraystretch{1.2}
\begin{table}[!htp] 
\caption{Residual effective cross-section (fB) at various cut flow stages for the
two described event topologies, as applied to the $M_{1/2} = 1000$~GeV \fsu5 benchmark and
the $\ttbar$ plus 2~Jets inclusive background at the LHC14.  The production level cross sections
and di-tau event selections are common to both scenarios.  Energies and momenta are in GeV.}
\label{tab:flow}
\begin{center}
\begin{tabular}{c c c c}
\hline 
Selection & $\ttbar + $Jets & Signal \\
\hline
Matched Production \quad & 645,000 & 192 \\
2 $\tau$'s, $P_{\rm T} > (40,20)$ \quad  & 2,230 & 10.1 \\
\hline
\multicolumn{3}{c}{\it Conventional Event Topology} \\
2 Jets, $P_{\rm T} > (400,200)$   \quad  & 15 & 4.2 \\
$\met > 700$    \quad   & 0.040 & 1.5 \\
\hline
\multicolumn{3}{c}{\it Boosted Event Topology} \\
2 Jets, $\Delta \eta > 3.5$ \& $P_{\rm T} > (75,50)$   \quad  & 88.1 & 0.78 \\
$\met > 700$    \quad   & 0.005 & 0.21 \\
\hline
\end{tabular}
\end{center}
\end{table}
\egroup

For the boosted event topology, increasing $P_{\rm T}$ of
the jets beyond the initial 20~GeV threshold is
actually quite detrimental, almost immediately.
Note in this case that the $\met$ contribution may be
balanced by more central non-tagged jets with much higher
$P_{\rm T}$.  However, we are concerned that such soft 
jets are not realistic for tagging in a high pile-up environment, and
instead adopt the more conservative bounds $P_{\rm T} > (75,50)$,
sacrificing approximately two thirds of the signal in the process.  The
background drops by a similar, if slightly larger, factor close to five.
Table~\ref{tab:flow} suggests that the hard jet cuts applied in the
conventional event topology remove quite a bit less signal and
quite a bit more background than the corresponding $\Delta \eta$ requirement imposed
in the boosted topology; however, subsequent application of an identical
$\met > 700$~GeV cut on missing transverse energy devastates the boosted
topology background component, while removing very little additional
signal by comparison.

After all cuts, the signal and background expectations
are both larger, by around 7 or 8 times, for the conventional event
topology, but the signal to background ratio S/B is essentially identical
for the two event topologies.  It should be emphasized that both scenarios,
and especially the boosted event topology, present such stringent suppression
of the $\ttbar + {\rm jets}$ background component, no more than about one
event per 25~${\rm fB}^{-1}$ of data, that the statistical and systematic
reliability of the analysis may become over stretched when extended to
the detailed comparison of two extremely small values; nevertheless, the central
message appears quite robust: backgrounds are extremely well controlled,
and signal has a realistic chance to present itself.

Fig.~(\ref{Fig:met_shape}) depicts the unity-normalized $\ttbar +$jets and \fsu5 event densities as a
function of missing transverse energy $\met$ for both the boosted and conventional event topologies,
with all other cuts applied.  The pair of distributions are found to be extremely similar in shape,
each exhibiting strong differentiation of signal and background for sufficiently strong $\met$ cuts.  The
heavier \fsu5 benchmarks extend more prevalently into the extremely large missing energy domain.  The
boosted event topology does demonstrate a somewhat harder background suppression slope in the distribution tail.
Moderate bin smoothing is applied here and in most subsequent plots.

Fig.~(\ref{Fig:met_sig}) exhibits the integrated signal significance as a function
of the missing transverse energy cut for the boosted and conventional event topologies.
The metric $S/\sqrt{1+B}$ is employed, comparing the count of signal events to the Gauss/Poisson fluctuation
inherent to the background estimation, with the numeral 1 employed as a low-background regulator.
A very mild scale dependence is observed, with lighter models favoring a bit less missing energy; this
asymmetry is strongly accentuated when employing the alternative significance metric $S/\sqrt{S+B}$,
because lighter models generate a large quantity of signal events, which may exceed or retain parity with
the background, such that significance scales with the signal square-root, disfavoring an over-strong cut.
Optimization with regard $S/\sqrt{1+B}$ is much less dependent upon the signal event scale, being
driven instead primarily by background elimination, which becomes particularly pronounced in the vicinity of $\met > 500 - 700$~GeV. 
The apparent tendency of the boosted event topology to favor a lighter $\met$ cut is attributable to the
previously described reduction in net event counts associated with this scenario, such that the regulator
term is dominant at moderate luminosities; nevertheless, the ratio S/B continues to benefit substantially from a harder cut.

\PlotSingle{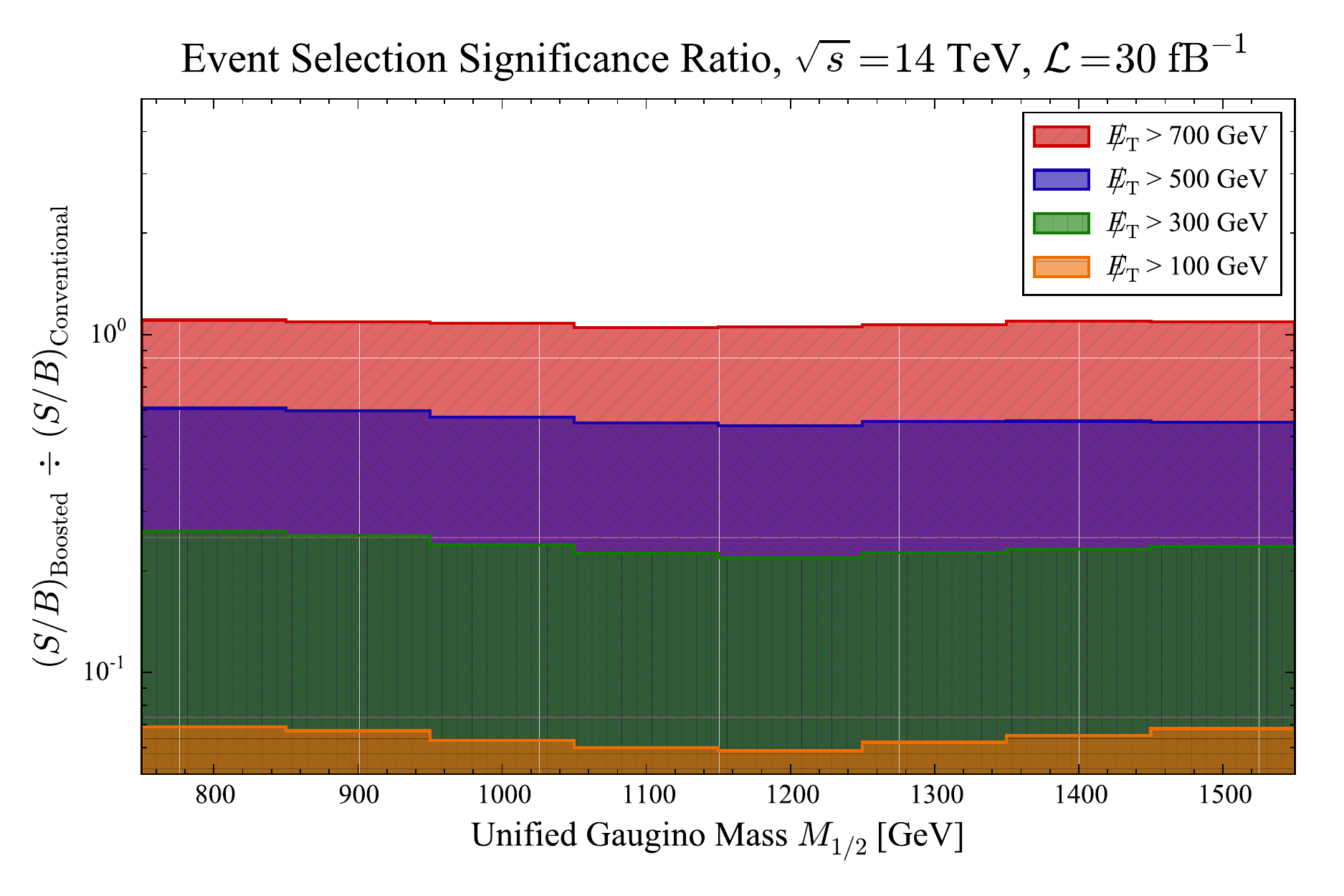}{3.4in}{Fig:sig_ratio}
{The ratio of boosted to conventional signal to background ratios is displayed as a function
of the gaugino scale parameter $M_{1/2}$ for various $\met$ cut thresholds.}

\PlotPairWide{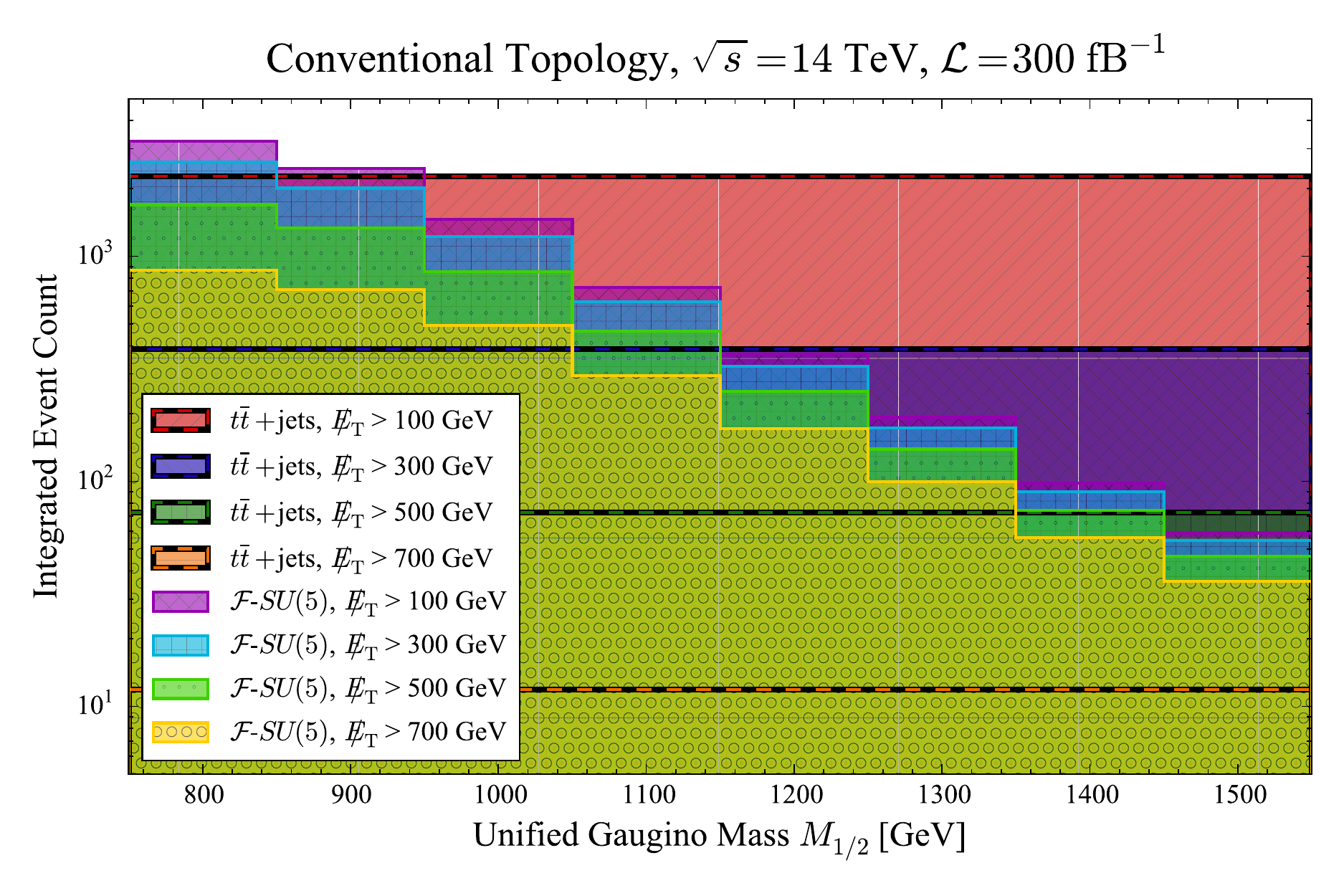}{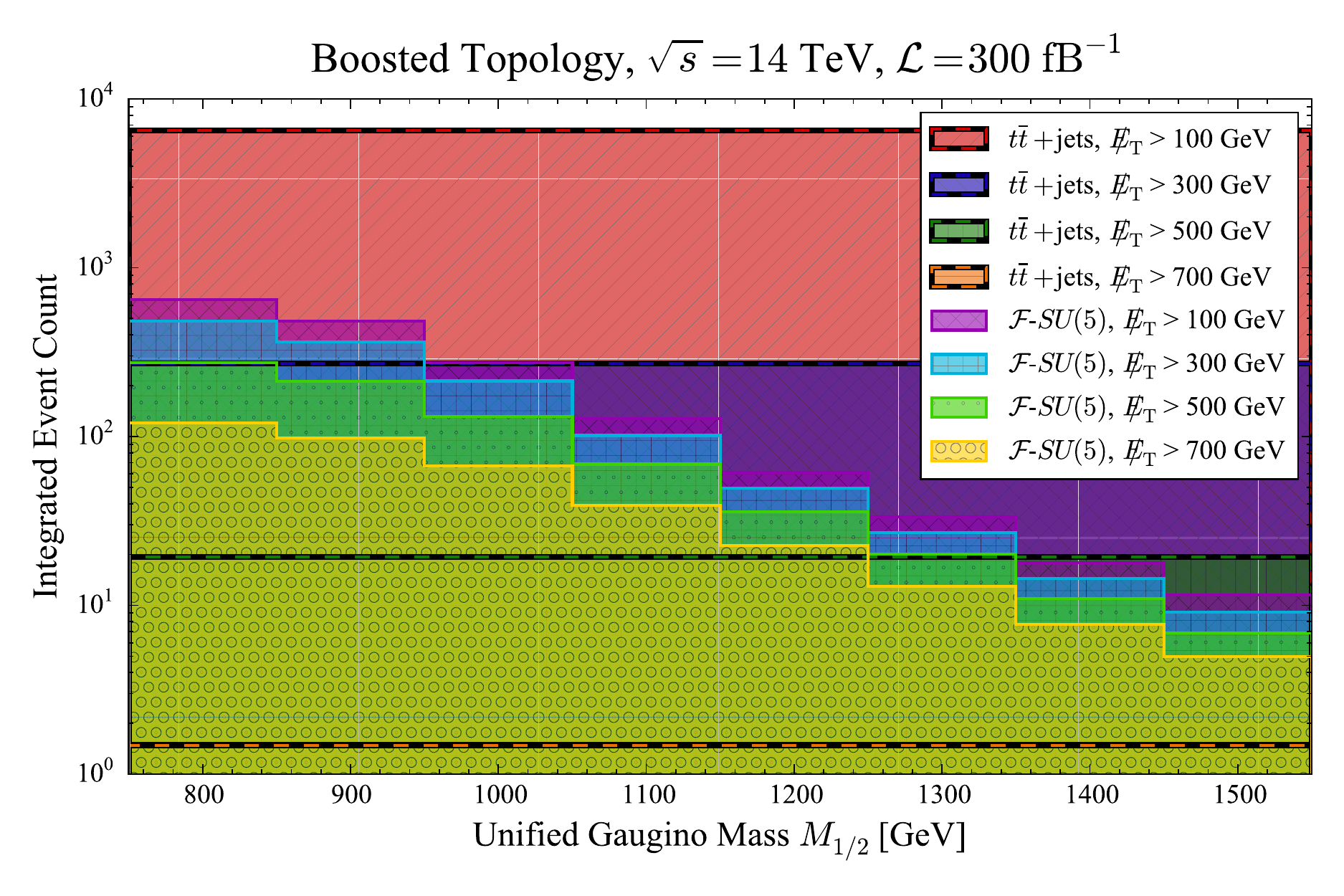}{Fig:event_counts}
{Expected \fsu5 signal and $\ttbar$ background counts at a luminosity of 300 events per femtobarn are depicted for
the boosted and conventional event topologies as a function of the gaugino scale parameter $M_{1/2}$ for various $\met$ cut thresholds.}

Fig.~(\ref{Fig:sig_ratio}) compares the boosted versus conventional signal to background ratios $S/B$ as a function of
the signal event scale for various common $\met$ cuts.  This ratio is observed 
to be comparable for both event topologies at a missing energy cut around 700~GeV, independent of the model scale.

Fig.~(\ref{Fig:event_counts}) depicts the raw \fsu5 signal and $\ttbar$ event counts for the boosted
and conventional event topologies as a function of the gaugino scale parameter $M_{1/2}$ for various
values of the missing energy cut at an integrated luminosity of 300 events per femtobarn.
A global suppression factor on the order of 10 is observed for application of the boosted event topology.
A dramatic discrimination is observed in the relative impact of background (much stronger) and signal (much milder) suppression with increasing missing energy. 
For $\met > 700$~GeV, the \fsu5 signal counts exceed the $\ttbar$ background counts for event scales across the viable model space,
up to $M_{1/2} = 1.5$~TeV and beyond.
A minimal signal count of more than 5 events is likewise maintained
over this full scale range, even for the boosted topology.  For a softer missing energy cut, e.g. $\met > 300$~GeV, the observed events
are expected to become background dominated at a much lighter scale, as low as $1.0$~TeV in this case.

\PlotPairWide{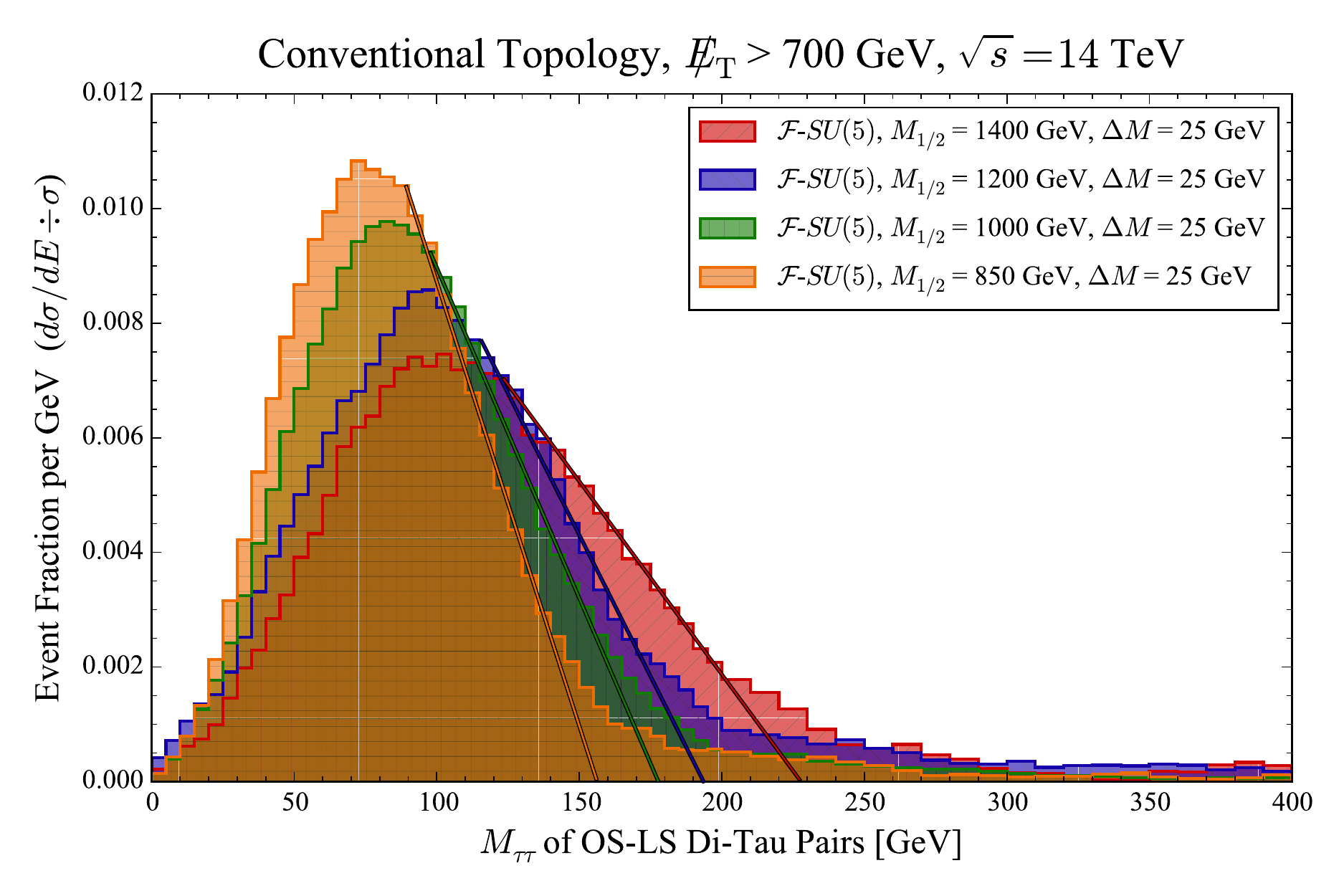}{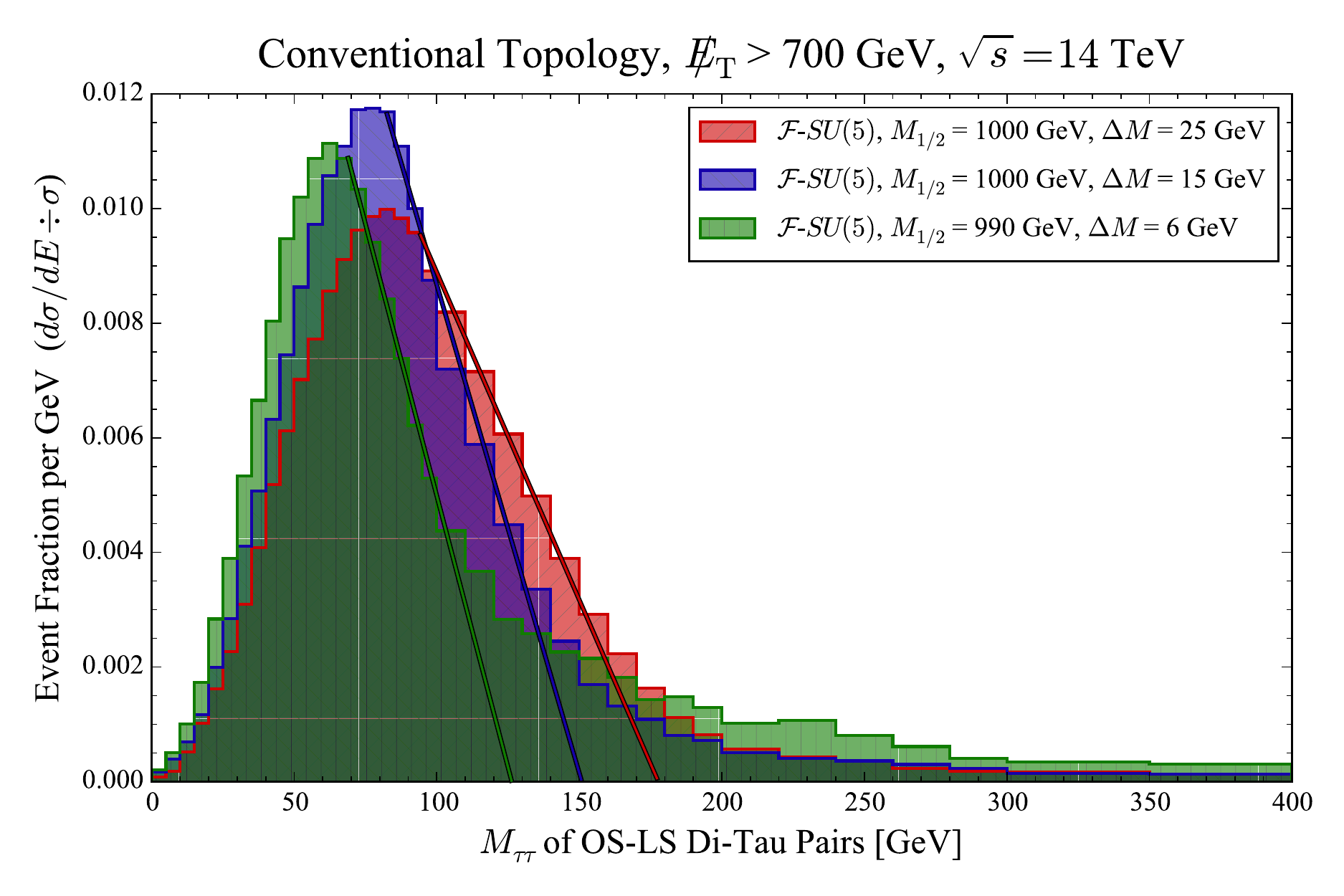}{Fig:mtt_shape}
{The conventional event topology opposite-sign minus like-sign di-tau invariant mass $M_{\tau\tau}$ distribution shape
is depicted for various gaugino mass $M_{1/2}$ scales and mass gaps
$\Delta M \equiv M_{{\tilde{\tau}}_{1}} - M_{{\tilde{\chi}}^0_1}$.}

\PlotPairWide{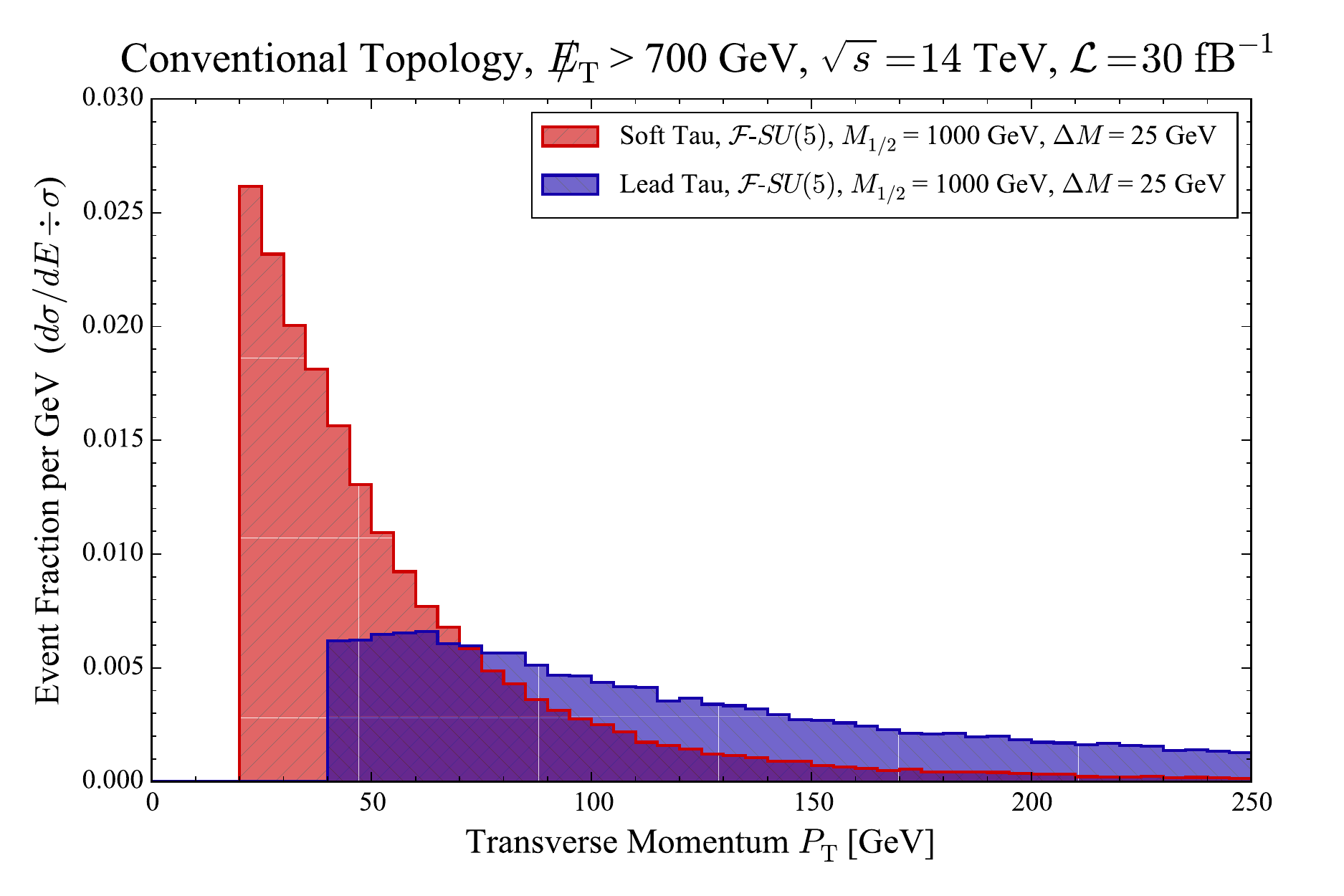}{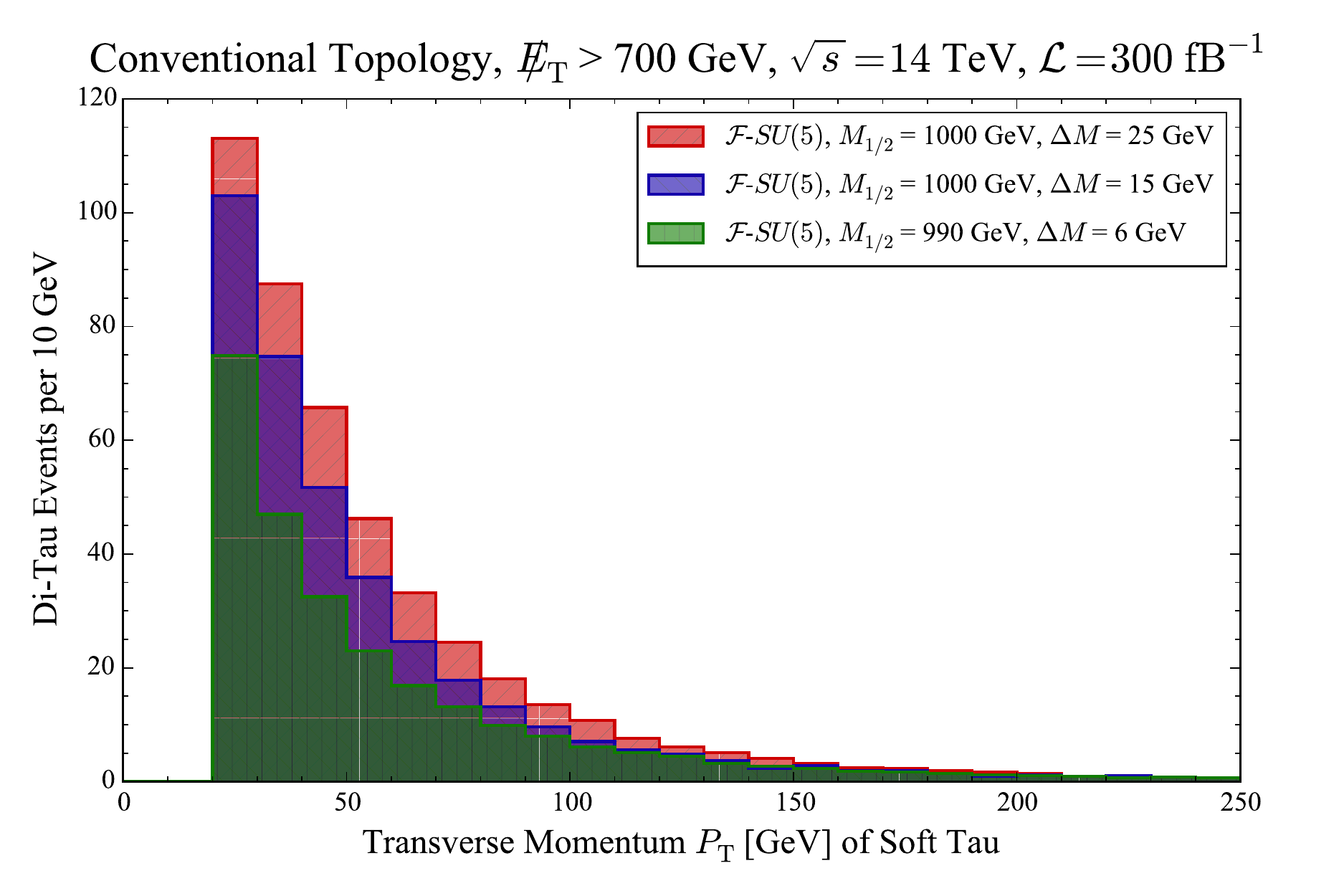}{Fig:pts}
{The distribution of transverse momenta $P_{\rm T}$ associated with individual hadronic tau elements is depicted
in terms of the luminosity-independent shape, and in terms of the absolute event yield at fixed
luminosity for various values of the mass gap $\Delta M \equiv M_{{\tilde{\tau}}_{1}} - M_{{\tilde{\chi}}^0_1}$.}

\section{The Neutralino-Stau System\label{sct:mtt}}

The structure of the neutralino-stau mass
hierarchy will be strongly imprinted upon the kinematic distribution of invariant masses $M_{\tau\tau}$
for opposite-sign (OS) di-tau pairs~\cite{Arnowitt:2008bz}.  After subtracting the density of like-sign (LS)
pairs, which serve as a statistical estimator for the density and shape of unassociated opposite sign production, the invariant mass cutoff
$M_{\tau\tau}^{\rm max}$~\cite{Arnowitt:2008bz} shown following may become visible, thereby providing a window
into the nature of dark matter and its potential compatibility with thermal processes~\cite{Arnowitt:2006jq,Dutta:2011kp}
that is largely orthogonal to other techniques.

\begin{equation}
M_{\tau\tau}^{\rm max} =
M_{{\tilde{\chi}}^0_2}
\sqrt{1-\frac{
M_{{\tilde{\tau}}_{1}}^2
}{
M_{{\tilde{\chi}}^0_2}^2
}}
\sqrt{1-\frac{
M_{{\tilde{\chi}}^0_1}^2
}{
M_{{\tilde{\tau}}_{1}}^2
}}
\label{eq:cutoff}
\end{equation}

Fig.~(\ref{Fig:mtt_shape}) depicts the opposite-sign (OS) minus like-sign (LS) di-tau invariant mass distribution shape
in the conventional event topology after applying a missing energy cut of $\met > 700$~GeV.  In the left-hand pane
the universal gaugino mass $M_{1/2}$ is varied as the stau-LSP mass gap $\Delta M \equiv M_{{\tilde{\tau}}_{1}} - M_{{\tilde{\chi}}^0_1}$
is held fixed at 25~GeV.  In the right-hand pane, the mass gap $\Delta M$ is fluctuated while holding the gaugino
mass $M_{1/2}$ (approximately) constant.  The distribution cutoff is visually extended in order to facilitate
a numerical comparison with theoretical expectations, as in Table~\ref{TAB:cutoff}.  Fluctuation of this extension
suggests a generic error on the order of 10~GeV in the limit of high statistics.  Variation of the bin sizing and
smoothing parameterization may induce a more or less comparably sized shift, to be combined in quadrature.

\bgroup
\def\arraystretch{1.2}
\begin{table}[!htp] 
\caption{Comparison between theoretical prediction and Monte-Carlo observation of the
opposite-sign minus like-sign di-tau invariant mass distribution cutoff for \fsu5
benchmarks with various gaugino mass $M_{1/2}$ and stau-LSP mass gap $\Delta M$ values.
The ``$M_{\tau\tau}^{\rm max}$'' value is computed by direct application of Eq.~(\ref{eq:cutoff}) to the associated spectrum.
The ``Visual'' entry references the extrapolated linear descents plotted in Fig.~(\ref{Fig:mtt_shape}), where applicable.
The ``$2\times$Peak'' value is established numerically, averaging over bin counts in
the upper third of the corresponding histogram plots (after moderate bin smoothing to wash out noise
peaks) and simply doubling the result. The ``With $\ttbar$\,'' column applies this same logic to the
\fsu5 signal and $\ttbar + {\rm jets}$~SM background combination.  A dash is presented when the signal
is highly unresolved relative to background. All entries are in units of GeV.}
\label{TAB:cutoff}
\begin{center}
\begin{tabular}{c c c c c c} 
\hline
~$M_{1/2}$~ & ~$\Delta M$~ & ~$M_{\tau\tau}^{\rm max}$~ & ~Visual~ & ~$2\times$Peak~ & ~With $\ttbar$~ \\
\hline
850 & 25 & 153 & 155 & 155 & 154 \\
1000 & 25 & 172 & 175 & 170 & 165 \\ 
1200 & 25 & 193 & 195 & 196 & 178 \\
1400 & 25 & 213 & 225 & 204 & - \\
\hline
821 & 15 & 123 & 135 & 135 & 139 \\
1000 & 15 & 140 & 150 & 155 & 150 \\
1400 & 16 & 172 & 175 & 183 & - \\
\hline
875 & 6 & 85 & 120 & 110 & 109 \\
990 & 6 & 94 & 125 & 125 & 115 \\
1400 & 6 & 111 & 155 & 145 & - \\
\hline
\end{tabular}
\end{center}
\end{table}
\egroup

Table~\ref{TAB:cutoff} examines the agreement between simulated data of a cascade-rich signal
and the theoretical prediction from Eq.~(\ref{eq:cutoff}) for the di-tau invariant mass cutoff $M_{\tau\tau}^{\rm max}$.
For a large mass gap $\Delta M \simeq 25$~GeV, which implies a non-thermal mechanism for dilution of the Bino
relic density, agreement is excellent.  Ability to discern a well correlated cutoff persists for the
background-polluted sample, although efficacy of the method is somewhat diminished in this case at heavier signal scales. 
Approaching the thermal mass gap around $\Delta M = 6$~GeV, where the soft tau element becomes increasingly difficult to resolve,
the cutoff extracted from simulation appears to systematically overestimate the Eq.~(\ref{eq:cutoff}) prediction. 

The tabulated elements, as well as Fig.~(\ref{Fig:mtt_shape}), use oversampled data.  In practice, the di-tau invariant
mass peak is a difficult measurement requiring high luminosity.  For the model samples under consideration,
very little is resolvable at a luminosity of $30$ fB$^{-1}$, but benchmarks in the vicinity of $M_{1/2} \sim 1$~TeV
present a sufficient event count for peak resolution at $300$ fB$^{-1}$.  Reduction of the $\met$ threshold substantially
improves event yields, but makes discrimination from the background very difficult. 

Fig.~(\ref{Fig:pts}) visually examines the distribution of transverse momenta $P_{\rm T}$ associated with individual hadronic
taus.  The left-hand pane plots the luminosity-independent shape of the soft and hard tau elements for $M_{1/2} = 1000$~GeV
and a stau-LSP mass gap $\Delta M = 25$~GeV.  The depicted shape is broadly representative of that observed in other relevant model scenarios.
A sharp decline is evident in the transverse momentum population of the softer tau, which is ostensibly associable with the phase space
constricted secondary decay $\tilde{\tau}_1 \to {\tilde{\chi}_1^0} + \tau$; this circumstance highlights the criticality of extending
detection acceptance for hadronic tau candidates to as low a transverse momentum threshold as is technically possible,
and at least to $P_{\rm T} \sim 20$~GeV.  The transverse momentum distribution of the lead tau, which is expected to arise
from the kinematically broad channel ${\tilde{\chi}_2^0} \to \tilde{\tau}_1 + \tau$, is rather flat; a harder cut here,
on the order that currently employed at $P_{\rm T} > 40$, does not appear detrimental.  The right-hand pane depicts the absolute distribution height
of the soft tau element at a luminosity of 300 events per femtobarn for various values of $\Delta M$.  As expected, this figure
suggests that it is substantially more difficult (though not impossible) to effectively probe di-tau production in the case of a narrow
stau-LSP mass gap; visibility of the daughter tau is more strongly dependent in this case upon a substantial upstream boosting of the parent stau.

\section{Discussion and Conclusions\label{sct:conclusions}}
In this paper we address a scenario where the gluino mass is in the 1 to 2 TeV range, whereas the first two squark masses and sleptons are heavy.
In particular, we have considered the No-Scale \fsu5 model, where the presence of additional vector like fields
may contribute to the generation of a Higgs mass near 126 GeV.
The light stop $\tilde{t}_1$ is lighter than the gluino in this model, which allows the gluino to decay on-shell into $\overline{t}\tilde{t}_1$.
The lighter stop may then decay directly into a top plus lightest neutralino,
or into a top plus (Wino-like) second lightest neutralino or bottom plus light chargino, which decay in turn to the lightest neutralino via stau.
The dark matter content may be obtained thermally by stau and (Bino-like) light neutralino coannihilation in the case of a small $\Delta M \simeq 6$~GeV.
The final state of such a cascade contains multiple $b$-quarks, $W$'s,  $\tau$ leptons and missing energy, and 
additional light flavor leptons, making $t\bar t$ the dominant background.

We first established  the scenario from the multi-$b$, leptons and $W$'s, so that the existence of third generation
can be surmised. We showed the signal can be differentiated from the leading background
(where we considered $t\bar t$ + multijet production with a 1\% $b$-fake rate)
by demanding multiple jets, multiple $b$-quarks, and multiple $W$'s in the final state,
in addition to very large missing transverse energy $\met$.
We calculated significances and showed that $M_{1/2}$ scales up to 1400 GeV ($M_{\tilde g} \sim 1900$~GeV) can be investigated
at the 14 TeV LHC with a 30 fB$^{-1}$ luminosity, requiring 8 jets with 2 heavy flavor tags and $\met > 700$~GeV.
In the multi-lepton cases, we showed that a strategy requiring at least 6 jets with a like sign dilepton, or four jets
with three leptons, can be utilized in conjunction with missing transverse energy to establish the model.
We showed that $M_{1/2}$ up to around 1200 GeV ($M_{\tilde g} \sim 1600$~GeV) can be investigated
in these channels at the 14 TeV LHC with a 30 fB$^{-1}$ luminosity; sensitivity may be extended to
$M_{1/2} \sim$ 1300-1400~GeV ($M_{\tilde g} \sim$ 1750-1900~GeV) with a 300 fB$^{-1}$ luminosity.
  
We investigated the coannihilation region by considering two analysis routes, each of which require
at least two $\tau$'s with $\pT > (40,20)$ GeV in $|\eta|<2.5$ and large missing energy.
In the ``Boosted Event Topology'', two tagging jets $j_{1,2}$ with $\pT > (75,50)$~GeV were required in $|\eta| \leq 5.0$  with 
an opposite hemisphere orientation $(\eta_1\times\eta_2 < 0)$, and an absolute separation in pseudo-rapidity of $|\eta_2-\eta_1|>3.5$.
In the ``Conventional Topology'', two leading jets $j_{1,2}$ in  $|\eta| < 2.5$
carry very large transverse momenta $\pT > (400,200)$~GeV.  To retain the cascade signal, no vetoes on heavy flavor jets
or light lepton flavors were enforced.  We demonstrated the cut flow optimization that leads to these selections,
and calculated the signal to background ratios S/B for both routes.
Both event topologies require a large $\met$ cut around 700 GeV to reduce the background.
After all cuts, the signal and background expectations
are both larger, by around 7 or 8 times, for the conventional event
topology, but the signal to background ratio S/B is essentially identical for the two event topologies.
With the optimized cuts, we investigated the  small $\Delta M$  region and found that
gaps of around 6 GeV can be probed for $M_{1/2} \sim 1000$ GeV
($M_{\tilde g} \sim 1300$~GeV) at $300$ fB$^{-1}$, although loss of the soft tau somewhat inhibits
resolution of the invariant $M_{\tau\tau}^{\rm max}$ mass
in this case; the experimental prospects are improved for larger $\Delta M$~GeV (around 15-25 GeV).

\section{Acknowledgments}

We would like to thank Yu Gao, Teruki Kamon and Nikolay Kolev for helpful discussions.
This work is supported in part by DOE Grant No.  DE-FG02-13ER42020, NSF Award PHY-1206044,
the Natural Science Foundation of China Grant Nos. 10821504, 11075194, 11135003, 11275246, and 11475238, the National Basic
Research Program of China (973 Program) under Grant No. 2010CB833000 (TL),
NASA Astrophysics Theory Grant NNH12ZDA001N (KS),
and the Sam Houston State University Enhancement Research Grant program (JWW).

\bibliography{bibliography}

\begin{thebibliography}{26}
\expandafter\ifx\csname natexlab\endcsname\relax\def\natexlab#1{#1}\fi
\expandafter\ifx\csname bibnamefont\endcsname\relax
  \def\bibnamefont#1{#1}\fi
\expandafter\ifx\csname bibfnamefont\endcsname\relax
  \def\bibfnamefont#1{#1}\fi
\expandafter\ifx\csname citenamefont\endcsname\relax
  \def\citenamefont#1{#1}\fi
\expandafter\ifx\csname url\endcsname\relax
  \def\url#1{\texttt{#1}}\fi
\expandafter\ifx\csname urlprefix\endcsname\relax\def\urlprefix{URL }\fi
\providecommand{\bibinfo}[2]{#2}
\providecommand{\eprint}[2][]{\url{#2}}

\bibitem[{\citenamefont{Aad et~al.}(2013)}]{Aad:2012fqa}
\bibinfo{author}{\bibfnamefont{G.}~\bibnamefont{Aad}} \bibnamefont{et~al.}
  (\bibinfo{collaboration}{ATLAS Collaboration}), {``}\bibinfo{title}{{Search
  for squarks and gluinos with the ATLAS detector in final states with jets and
  missing transverse momentum using 4.7 fb$^{-1}$ of $\sqrt{s}=7$ TeV
  proton-proton collision data}},{''} \bibinfo{journal}{Phys.Rev.}
  \textbf{\bibinfo{volume}{D87}}, \bibinfo{pages}{012008}
  (\bibinfo{year}{2013}), \eprint{1208.0949}.

\bibitem[{\citenamefont{Aad et~al.}(2012)}]{Aad:2012hm}
\bibinfo{author}{\bibfnamefont{G.}~\bibnamefont{Aad}} \bibnamefont{et~al.}
  (\bibinfo{collaboration}{ATLAS Collaboration}), {``}\bibinfo{title}{{Hunt for
  new phenomena using large jet multiplicities and missing transverse momentum
  with ATLAS in ${\rm fb^{-1}}$ of $\sqrt{s}$ = 7 TeV proton-proton
  collisions}},{''} (\bibinfo{year}{2012}), \eprint{1206.1760}.

\bibitem[{\citenamefont{Chatrchyan et~al.}(2012)}]{Chatrchyan:2012lia}
\bibinfo{author}{\bibfnamefont{S.}~\bibnamefont{Chatrchyan}}
  \bibnamefont{et~al.} (\bibinfo{collaboration}{CMS Collaboration}),
  {``}\bibinfo{title}{{Search for new physics in the multijet and missing
  transverse momentum final state in proton-proton collisions at $\sqrt{s} = 7$
  TeV}},{''} \bibinfo{journal}{Phys.Rev.Lett.} \textbf{\bibinfo{volume}{109}},
  \bibinfo{pages}{171803} (\bibinfo{year}{2012}), \eprint{1207.1898}.

\bibitem[{\citenamefont{collaboration}(2013)}]{TheATLAScollaboration:2013fha}
\bibinfo{author}{\bibfnamefont{T.~A.} \bibnamefont{collaboration}},
  {``}\bibinfo{title}{{Search for squarks and gluinos with the ATLAS detector
  in final states with jets and missing transverse momentum and 20.3 fb$^{-1}$
  of $\sqrt{s}=8$ TeV proton-proton collision data}},{''}
  (\bibinfo{year}{2013}).

\bibitem[{\citenamefont{Khachatryan et~al.}(2014{\natexlab{a}})}]{CMS:2014dpa}
\bibinfo{author}{\bibfnamefont{V.}~\bibnamefont{Khachatryan}}
  \bibnamefont{et~al.} (\bibinfo{collaboration}{CMS Collaboration}),
  {``}\bibinfo{title}{{Searches for supersymmetry based on events with b jets
  and four W bosons in pp collisions at 8 TeV}},{''}
  (\bibinfo{year}{2014}{\natexlab{a}}), \eprint{1412.4109}.

\bibitem[{\citenamefont{Li et~al.}(2013)\citenamefont{Li, Maxin, Nanopoulos,
  and Walker}}]{Li:2013naa}
\bibinfo{author}{\bibfnamefont{T.}~\bibnamefont{Li}},
  \bibinfo{author}{\bibfnamefont{J.~A.} \bibnamefont{Maxin}},
  \bibinfo{author}{\bibfnamefont{D.~V.} \bibnamefont{Nanopoulos}},
  \bibnamefont{and} \bibinfo{author}{\bibfnamefont{J.~W.}
  \bibnamefont{Walker}}, {``}\bibinfo{title}{{No-Scale ${\cal F}$-$SU(5)$ in
  the Light of LHC, Planck and XENON}},{''} \bibinfo{journal}{Jour.Phys.}
  \textbf{\bibinfo{volume}{G40}}, \bibinfo{pages}{115002}
  (\bibinfo{year}{2013}), \eprint{1305.1846}.

\bibitem[{\citenamefont{Li et~al.}(2011)\citenamefont{Li, Maxin, Nanopoulos,
  and Walker}}]{Maxin:2011hy}
\bibinfo{author}{\bibfnamefont{T.}~\bibnamefont{Li}},
  \bibinfo{author}{\bibfnamefont{J.~A.} \bibnamefont{Maxin}},
  \bibinfo{author}{\bibfnamefont{D.~V.} \bibnamefont{Nanopoulos}},
  \bibnamefont{and} \bibinfo{author}{\bibfnamefont{J.~W.}
  \bibnamefont{Walker}}, {``}\bibinfo{title}{{The Ultrahigh jet multiplicity
  signal of stringy no-scale ${\cal F}$-$SU(5)$ at the $\sqrt{s}= 7$ TeV
  LHC}},{''} \bibinfo{journal}{Phys.Rev.} \textbf{\bibinfo{volume}{D84}},
  \bibinfo{pages}{076003} (\bibinfo{year}{2011}), \eprint{1103.4160}.

\bibitem[{\citenamefont{Li et~al.}(2012)\citenamefont{Li, Maxin, Nanopoulos,
  and Walker}}]{Li:2011ab}
\bibinfo{author}{\bibfnamefont{T.}~\bibnamefont{Li}},
  \bibinfo{author}{\bibfnamefont{J.~A.} \bibnamefont{Maxin}},
  \bibinfo{author}{\bibfnamefont{D.~V.} \bibnamefont{Nanopoulos}},
  \bibnamefont{and} \bibinfo{author}{\bibfnamefont{J.~W.}
  \bibnamefont{Walker}}, {``}\bibinfo{title}{{A Higgs Mass Shift to 125 GeV and
  A Multi-Jet Supersymmetry Signal: Miracle of the Flippons at the $\sqrt{s} =
  7$~TeV LHC}},{''} \bibinfo{journal}{Phys.Lett.}
  \textbf{\bibinfo{volume}{B710}}, \bibinfo{pages}{207} (\bibinfo{year}{2012}),
  \eprint{1112.3024}.

\bibitem[{\citenamefont{Aparicio et~al.}(2014)\citenamefont{Aparicio, Cicoli,
  Krippendorf, Maharana, Muia et~al.}}]{Aparicio:2014wxa}
\bibinfo{author}{\bibfnamefont{L.}~\bibnamefont{Aparicio}},
  \bibinfo{author}{\bibfnamefont{M.}~\bibnamefont{Cicoli}},
  \bibinfo{author}{\bibfnamefont{S.}~\bibnamefont{Krippendorf}},
  \bibinfo{author}{\bibfnamefont{A.}~\bibnamefont{Maharana}},
  \bibinfo{author}{\bibfnamefont{F.}~\bibnamefont{Muia}}, \bibnamefont{et~al.},
  {``}\bibinfo{title}{{Sequestered de Sitter String Scenarios:
  Soft-terms}},{''} \bibinfo{journal}{JHEP} \textbf{\bibinfo{volume}{1411}},
  \bibinfo{pages}{071} (\bibinfo{year}{2014}), \eprint{1409.1931}.

\bibitem[{\citenamefont{Dutta et~al.}(2014)\citenamefont{Dutta, Flanagan,
  Gurrola, Johns, Kamon et~al.}}]{Dutta:2013gga}
\bibinfo{author}{\bibfnamefont{B.}~\bibnamefont{Dutta}},
  \bibinfo{author}{\bibfnamefont{W.}~\bibnamefont{Flanagan}},
  \bibinfo{author}{\bibfnamefont{A.}~\bibnamefont{Gurrola}},
  \bibinfo{author}{\bibfnamefont{W.}~\bibnamefont{Johns}},
  \bibinfo{author}{\bibfnamefont{T.}~\bibnamefont{Kamon}},
  \bibnamefont{et~al.}, {``}\bibinfo{title}{{Probing Compressed Top Squarks at
  the LHC at 14 TeV}},{''} \bibinfo{journal}{Phys.Rev.}
  \textbf{\bibinfo{volume}{D90}}, \bibinfo{pages}{095022}
  (\bibinfo{year}{2014}), \eprint{1312.1348}.

\bibitem[{\citenamefont{Arnowitt et~al.}(2008)\citenamefont{Arnowitt, Dutta,
  Gurrola, Kamon, Krislock et~al.}}]{Arnowitt:2008bz}
\bibinfo{author}{\bibfnamefont{R.~L.} \bibnamefont{Arnowitt}},
  \bibinfo{author}{\bibfnamefont{B.}~\bibnamefont{Dutta}},
  \bibinfo{author}{\bibfnamefont{A.}~\bibnamefont{Gurrola}},
  \bibinfo{author}{\bibfnamefont{T.}~\bibnamefont{Kamon}},
  \bibinfo{author}{\bibfnamefont{A.}~\bibnamefont{Krislock}},
  \bibnamefont{et~al.}, {``}\bibinfo{title}{{Determining the Dark Matter Relic
  Density in the mSUGRA (~X0(1))-~tau Co-Annhiliation Region at the LHC}},{''}
  \bibinfo{journal}{Phys. Rev. Lett.} \textbf{\bibinfo{volume}{100}},
  \bibinfo{pages}{231802} (\bibinfo{year}{2008}), \eprint{0802.2968}.

\bibitem[{\citenamefont{Ellis et~al.}(2002)\citenamefont{Ellis, Nanopoulos, and
  Olive}}]{Ellis:2001kg}
\bibinfo{author}{\bibfnamefont{J.~R.} \bibnamefont{Ellis}},
  \bibinfo{author}{\bibfnamefont{D.~V.} \bibnamefont{Nanopoulos}},
  \bibnamefont{and} \bibinfo{author}{\bibfnamefont{K.~A.} \bibnamefont{Olive}},
  {``}\bibinfo{title}{{Lower limits on soft supersymmetry breaking scalar
  masses}},{''} \bibinfo{journal}{Phys.\ Lett.} \textbf{\bibinfo{volume}{{\bf
  B525}}}, \bibinfo{pages}{308} (\bibinfo{year}{2002}), \eprint{arXiv:0109288}.

\bibitem[{\citenamefont{Ellis et~al.}(2010)\citenamefont{Ellis, Mustafayev, and
  Olive}}]{Ellis:2010jb}
\bibinfo{author}{\bibfnamefont{J.}~\bibnamefont{Ellis}},
  \bibinfo{author}{\bibfnamefont{A.}~\bibnamefont{Mustafayev}},
  \bibnamefont{and} \bibinfo{author}{\bibfnamefont{K.~A.} \bibnamefont{Olive}},
  {``}\bibinfo{title}{{Resurrecting No-Scale Supergravity Phenomenology}},{''}
  \bibinfo{journal}{Eur. Phys. J.} \textbf{\bibinfo{volume}{C69}},
  \bibinfo{pages}{219} (\bibinfo{year}{2010}), \eprint{1004.5399}.

\bibitem[{\citenamefont{Djouadi et~al.}(2007)\citenamefont{Djouadi, Kneur, and
  Moultaka}}]{Djouadi:2002ze}
\bibinfo{author}{\bibfnamefont{A.}~\bibnamefont{Djouadi}},
  \bibinfo{author}{\bibfnamefont{J.-L.} \bibnamefont{Kneur}}, \bibnamefont{and}
  \bibinfo{author}{\bibfnamefont{G.}~\bibnamefont{Moultaka}},
  {``}\bibinfo{title}{{SuSpect: A Fortran code for the supersymmetric and Higgs
  particle spectrum in the MSSM}},{''} \bibinfo{journal}{Comput. Phys. Commun.}
  \textbf{\bibinfo{volume}{176}}, \bibinfo{pages}{426} (\bibinfo{year}{2007}),
  \eprint{hep-ph/0211331}.

\bibitem[{\citenamefont{Alwall et~al.}(2011)\citenamefont{Alwall, Herquet,
  Maltoni, Mattelaer, and Stelzer}}]{Alwall:2011uj}
\bibinfo{author}{\bibfnamefont{J.}~\bibnamefont{Alwall}},
  \bibinfo{author}{\bibfnamefont{M.}~\bibnamefont{Herquet}},
  \bibinfo{author}{\bibfnamefont{F.}~\bibnamefont{Maltoni}},
  \bibinfo{author}{\bibfnamefont{O.}~\bibnamefont{Mattelaer}},
  \bibnamefont{and} \bibinfo{author}{\bibfnamefont{T.}~\bibnamefont{Stelzer}},
  {``}\bibinfo{title}{{MadGraph 5 : Going Beyond}},{''} \bibinfo{journal}{JHEP}
  \textbf{\bibinfo{volume}{1106}}, \bibinfo{pages}{128} (\bibinfo{year}{2011}),
  \eprint{1106.0522}.

\bibitem[{\citenamefont{Sjostrand et~al.}(2006)\citenamefont{Sjostrand, Mrenna,
  and Skands}}]{Sjostrand:2006za}
\bibinfo{author}{\bibfnamefont{T.}~\bibnamefont{Sjostrand}},
  \bibinfo{author}{\bibfnamefont{S.}~\bibnamefont{Mrenna}}, \bibnamefont{and}
  \bibinfo{author}{\bibfnamefont{P.~Z.} \bibnamefont{Skands}},
  {``}\bibinfo{title}{{PYTHIA 6.4 Physics and Manual}},{''}
  \bibinfo{journal}{JHEP} \textbf{\bibinfo{volume}{05}}, \bibinfo{pages}{026}
  (\bibinfo{year}{2006}), \eprint{hep-ph/0603175}.

\bibitem[{\citenamefont{Conway et~al.}(2009)}]{PGS4}
\bibinfo{author}{\bibfnamefont{J.}~\bibnamefont{Conway}} \bibnamefont{et~al.},
  {``}\bibinfo{title}{PGS4: Pretty Good (Detector) Simulation},{''}
  (\bibinfo{year}{2009}),
  \urlprefix\url{http://www.physics.ucdavis.edu/~conway/research/}.

\bibitem[{\citenamefont{Walker}(2012)}]{Walker:2012vf}
\bibinfo{author}{\bibfnamefont{J.~W.} \bibnamefont{Walker}},
  {``}\bibinfo{title}{{CutLHCO: A Consumer-Level Tool for Implementing Generic
  Collider Data Selection Cuts in the Search for New Physics}},{''}
  (\bibinfo{year}{2012}), \eprint{1207.3383}.

\bibitem[{\citenamefont{Walker}(2014)}]{aeacus}
\bibinfo{author}{\bibfnamefont{J.~W.} \bibnamefont{Walker}},
  {``}\bibinfo{title}{AEACuS 3.4},{''} (\bibinfo{year}{2014}),
  \urlprefix\url{http://www.joelwalker.net/code/aeacus.tar.gz}.

\bibitem[{\citenamefont{Khachatryan
  et~al.}(2014{\natexlab{b}})}]{Khachatryan:2014gha}
\bibinfo{author}{\bibfnamefont{V.}~\bibnamefont{Khachatryan}}
  \bibnamefont{et~al.} (\bibinfo{collaboration}{CMS Collaboration}),
  {``}\bibinfo{title}{{Search for massive resonances decaying into pairs of
  boosted bosons in semi-leptonic final states at $\sqrt{s} =$ 8 TeV}},{''}
  \bibinfo{journal}{JHEP} \textbf{\bibinfo{volume}{1408}}, \bibinfo{pages}{174}
  (\bibinfo{year}{2014}{\natexlab{b}}), \eprint{1405.3447}.

\bibitem[{\citenamefont{Khachatryan
  et~al.}(2014{\natexlab{c}})}]{Khachatryan:2014hpa}
\bibinfo{author}{\bibfnamefont{V.}~\bibnamefont{Khachatryan}}
  \bibnamefont{et~al.} (\bibinfo{collaboration}{CMS Collaboration}),
  {``}\bibinfo{title}{{Search for massive resonances in dijet systems
  containing jets tagged as W or Z boson decays in pp collisions at $ \sqrt{s}
  $ = 8 TeV}},{''} \bibinfo{journal}{JHEP} \textbf{\bibinfo{volume}{1408}},
  \bibinfo{pages}{173} (\bibinfo{year}{2014}{\natexlab{c}}),
  \eprint{1405.1994}.

\bibitem[{\citenamefont{Dutta et~al.}(2013)\citenamefont{Dutta, Gurrola, Johns,
  Kamon, Sheldon et~al.}}]{Dutta:2012xe}
\bibinfo{author}{\bibfnamefont{B.}~\bibnamefont{Dutta}},
  \bibinfo{author}{\bibfnamefont{A.}~\bibnamefont{Gurrola}},
  \bibinfo{author}{\bibfnamefont{W.}~\bibnamefont{Johns}},
  \bibinfo{author}{\bibfnamefont{T.}~\bibnamefont{Kamon}},
  \bibinfo{author}{\bibfnamefont{P.}~\bibnamefont{Sheldon}},
  \bibnamefont{et~al.}, {``}\bibinfo{title}{{Vector Boson Fusion Processes as a
  Probe of Supersymmetric Electroweak Sectors at the LHC}},{''}
  \bibinfo{journal}{Phys.Rev.} \textbf{\bibinfo{volume}{D87}},
  \bibinfo{pages}{035029} (\bibinfo{year}{2013}), \eprint{1210.0964}.

\bibitem[{\citenamefont{Delannoy
  et~al.}(2013{\natexlab{a}})\citenamefont{Delannoy, Dutta, Gurrola, Johns,
  Kamon et~al.}}]{Delannoy:2013ata}
\bibinfo{author}{\bibfnamefont{A.~G.} \bibnamefont{Delannoy}},
  \bibinfo{author}{\bibfnamefont{B.}~\bibnamefont{Dutta}},
  \bibinfo{author}{\bibfnamefont{A.}~\bibnamefont{Gurrola}},
  \bibinfo{author}{\bibfnamefont{W.}~\bibnamefont{Johns}},
  \bibinfo{author}{\bibfnamefont{T.}~\bibnamefont{Kamon}},
  \bibnamefont{et~al.}, {``}\bibinfo{title}{{Probing Dark Matter at the LHC
  using Vector Boson Fusion Processes}},{''} \bibinfo{journal}{Phys.Rev.Lett.}
  \textbf{\bibinfo{volume}{111}}, \bibinfo{pages}{061801}
  (\bibinfo{year}{2013}{\natexlab{a}}), \eprint{1304.7779}.

\bibitem[{\citenamefont{Delannoy
  et~al.}(2013{\natexlab{b}})\citenamefont{Delannoy, Dutta, Gurrola, Johns,
  Kamon et~al.}}]{Delannoy:2013dla}
\bibinfo{author}{\bibfnamefont{A.~G.} \bibnamefont{Delannoy}},
  \bibinfo{author}{\bibfnamefont{B.}~\bibnamefont{Dutta}},
  \bibinfo{author}{\bibfnamefont{A.}~\bibnamefont{Gurrola}},
  \bibinfo{author}{\bibfnamefont{W.}~\bibnamefont{Johns}},
  \bibinfo{author}{\bibfnamefont{T.}~\bibnamefont{Kamon}},
  \bibnamefont{et~al.}, {``}\bibinfo{title}{{Probing Supersymmetric Dark Matter
  and the Electroweak Sector using Vector Boson Fusion Processes: A Snowmass
  Whitepaper}},{''} (\bibinfo{year}{2013}{\natexlab{b}}), \eprint{1308.0355}.

\bibitem[{\citenamefont{Arnowitt et~al.}(2006)\citenamefont{Arnowitt, Dutta,
  Kamon, Kolev, and Toback}}]{Arnowitt:2006jq}
\bibinfo{author}{\bibfnamefont{R.~L.} \bibnamefont{Arnowitt}},
  \bibinfo{author}{\bibfnamefont{B.}~\bibnamefont{Dutta}},
  \bibinfo{author}{\bibfnamefont{T.}~\bibnamefont{Kamon}},
  \bibinfo{author}{\bibfnamefont{N.}~\bibnamefont{Kolev}}, \bibnamefont{and}
  \bibinfo{author}{\bibfnamefont{D.~A.} \bibnamefont{Toback}},
  {``}\bibinfo{title}{{Detection of SUSY in the stau-neutralino coannihilation
  region at the LHC}},{''} \bibinfo{journal}{Phys.Lett.}
  \textbf{\bibinfo{volume}{B639}}, \bibinfo{pages}{46} (\bibinfo{year}{2006}),
  \eprint{hep-ph/0603128}.

\bibitem[{\citenamefont{Dutta et~al.}(2012)\citenamefont{Dutta, Kamon,
  Krislock, Sinha, and Wang}}]{Dutta:2011kp}
\bibinfo{author}{\bibfnamefont{B.}~\bibnamefont{Dutta}},
  \bibinfo{author}{\bibfnamefont{T.}~\bibnamefont{Kamon}},
  \bibinfo{author}{\bibfnamefont{A.}~\bibnamefont{Krislock}},
  \bibinfo{author}{\bibfnamefont{K.}~\bibnamefont{Sinha}}, \bibnamefont{and}
  \bibinfo{author}{\bibfnamefont{K.}~\bibnamefont{Wang}},
  {``}\bibinfo{title}{{Diagnosis of Supersymmetry Breaking Mediation Schemes by
  Mass Reconstruction at the LHC}},{''} \bibinfo{journal}{Phys.Rev.}
  \textbf{\bibinfo{volume}{D85}}, \bibinfo{pages}{115007}
  (\bibinfo{year}{2012}), \eprint{1112.3966}.

\end{thebibliography}

\end{document}